\documentclass[aps,prb,twocolumn,showpacs,superscriptaddress,10pt,longbibliography]{revtex4-2}
\usepackage[colorlinks = true,linkcolor = red,citecolor = magenta]{hyperref}
\usepackage[sort&compress]{natbib}
\usepackage{scalerel}
\usepackage[normalem]{ulem}
\usepackage{xcolor}
\usepackage{bm}
\usepackage{amsmath}
\usepackage{amsfonts}
\usepackage{amssymb}
\usepackage{graphicx}
\usepackage{subfigure}
\usepackage{mathtools}
\usepackage{physics}
\usepackage{dsfont}
\usepackage{float}
\usepackage{lmodern}
\usepackage[many]{tcolorbox}
\usepackage{empheq}
\usepackage{cleveref}
\usepackage{textcomp}
\usepackage{soul}
\usepackage{wasysym}

\setcitestyle{citesep={,\kern-0.25em}}

\usepackage{tikz}
\definecolor{lime}{HTML}{A6CE39}
\DeclareRobustCommand{\orcidicon}{%
    \begin{tikzpicture}
    \draw[lime, fill=lime] (0,0) 
    circle [radius=0.13] 
    node[white] {{\fontfamily{qag}\selectfont\tiny ID}};
    \draw[white, fill=white] (-0.0625,0.095) 
    circle [radius=0.007];
    \end{tikzpicture}
    \hspace{-2mm}}

\newcommand{\orcidAM}{\href{https://orcid.org/0000-0002-7307-5922}{\orcidicon}}
\newcommand{\orcidJK}{\href{https://orcid.org/0000-0003-0998-9460}{\orcidicon}}

\newcommand{\mh}{\mathcal{H}}
\newcommand{\md}{\mathcal{D}}

\usepackage{color}

\begin{document}
\title{Anomalous localization in spin chains with tilted interactions}

\author{Arindam Mallick\orcidAM}
\email{arindam.mallick@uj.edu.pl}
\affiliation{Instytut Fizyki Teoretycznej, Uniwersytet Jagielloński, Łojasiewicza 11, 30-348 Kraków, Poland}

\author{Jakub Zakrzewski\orcidJK}
\email{jakub.zakrzewski@uj.edu.pl}
\affiliation{Instytut Fizyki Teoretycznej, Uniwersytet Jagielloński, Łojasiewicza 11, 30-348 Kraków, Poland}
\affiliation{Mark Kac Complex Systems Research Center, Uniwersytet Jagielloński, 30-348 Kraków, Poland}

\date{\today}

\begin{abstract}
Quantum simulators of lattice gauge theories involve dynamics of typically short-ranged interacting particles and dynamical fields. Elimination of the latter via Gauss law leads to infinite range interactions as exemplified by the  Schwinger model in a staggered formalism. This motivates the study of long-range interactions, not necessarily diminishing with the distance.  Here we consider localization properties of a spin chain with interaction strength growing linearly along the chain as for the Schwinger model. We generalize the problem to models with different interaction ranges. Using exact diagonalization we find the participation ratio of all eigenstates, which allows us to quantify the localization volume in  Hilbert space. Surprisingly, the localization volume changes nonmonotonically with the interaction range. Our study is relevant for quantum simulators of lattice gauge theories implemented in state-of-the-art cold atom/ion devices, and it could help to reveal hidden features in disorder-free confinement phenomena in long-range interacting systems.
  \end{abstract}

\maketitle

\section{Introduction}

Localization, or more generally non-ergodicity, for interacting many-particle systems has been a subject of intensive research under the umbrella term many-body localization (MBL) following the seminal works in Refs.~\cite{Gornyi05,basko2006metal} (for reviews, see \cite{abanin2019colloquium, Alet18, Sierant24}). MBL breaks the conventional understanding of thermalization as expressed via the Eigenstate Thermalization Hypothesis (ETH) \cite{Peres84,Deutsch91,Srednicki94}. While nonergodic dynamics have been associated with the presence of disorder in the system, either fully random \cite{Gornyi05, basko2006metal, oganesyan2007localization} or quasi-periodic \cite{iyer2013many, Adith21},  recently localization without disorder has been drawing a lot of attention. The latter happens in the presence of energy constraints, e.g., linear tilt (dc field) or harmonic traps \cite{nieuwenburg2019bloch, schulz2019stark, Taylor20, Guo20, Scherg21, doggen2021stark, mallick2021wannier, yao2021nonergodic, mallick2022correlated, Lydzba24,Nandy24}; 
gauge-field-induced confinement \cite{kormos2016realtime, chanda2020confinement, Yang2020hilbert, Iadecola19, Iadecola19a, Surace20, Magnifico20}; in the presence of other species of particles that effectively induce random environment \cite{Gavish05,smith2017disorder}; in presence of flatbands \cite{danieli2020manybody, mallick2021wannier, mallick2022correlated}; etc. The study of long-range, e.g., dipolar interaction effects on such models also has a rich history \cite{Burin15, Burin15b, MaksymovBurin20, Li21, Adith23, Korbmacher23, cheng2023manybody, jiang2023stark}, with recent progress being related to experimental progress in the cooling and trapping of dipolar matter \cite{Baier16, Su23, Chomaz23}. 

Simulations of lattice gauge theories in different cold-atom settings have developed rapidly recently \cite{Wiese13,Banuls20,Aidelsburger21} due to progress in various experimental platforms \cite{Martinez16,Schweizer19,Mil20,Yang20}. In many of the proposed implementations, the particles reside in the sites of the lattice while fields live on its bonds \cite{Wiese13}. Such a model is typically assumed to have short-range interactions among particles {and dynamical gauge fields}. The presence of the local Gauss law provides a restriction on particles and fields and is used to eliminate the dynamical fields. The price to pay is that the resulting interaction among the particles becomes infinitely long-ranged (see the Appendix \ref{app:schw} for an example of the Schwinger model). These dynamics may have unexpected properties if viewed from a general point of interacting many-body physics. Here, in the absence of disorder, one expects, on the basis of ETH \cite{Deutsch91,Srednicki94}, the dominance of ergodic motion even for short-range interactions. The longer range should lead to even faster ergodization. Instead, for the lattice gauge models one often observes nonergodic dynamics resulting, e.g., in the appearance of confinement \cite{kormos2016realtime, chanda2020confinement, Yang2020hilbert, Iadecola19, Iadecola19a, Surace20, Magnifico20} as well as the existence of unusually regular states embedded in otherwise ergodic typical eigenstates---the so-called many-body quantum scars \cite{Bernien17,Iadecola19, Iadecola19a,Adith22,Szoldra22}.

Inspired by these developments, we introduce in this article a one-dimensional (1D) spin chain  model in which the interaction 
strength is inhomogeneous and, in particular, increases with the distance along the lattice. While typically the interaction strength decays as a power law or exponentially with the distance \cite{Korbmacher23, defenu2023longrange}, we consider the case in which it grows linearly, not only with respect to the relative distance but also with the explicit location of spins. In that way, it also differs from the 1D Coulomb model in the continuum where interaction strength grows with the relative distance between charge densities only \cite{nandkishore2017manybody}.

The model is motivated by lattice gauge theory and in particular the implementation of the Schwinger model on the lattice \cite{kogut1975hamiltonian, banks1976strong}. As described in detail in Appendix \ref{app:schw}, this connection appears when the range of interactions matches the system size. In the other extreme, the model with nearest-neighbor interactions (dependent, however, on the position) may be realized for spinless fermions by appropriately adapting the interactions via, e.g., a Feshbach resonance controlled magnetic field \cite{chin2010feshbach}. A similar model with nearest neighbor linearly growing interaction has been studied recently \cite{wang2024eigenstate}. We consider below the whole family of models with different interaction ranges.

By numerical full diagonalization using the Quspin python library \cite{Weinberg17, Weinberg19}, we study the localization volume, i.e., the participation ratio (PR) in Hilbert space spanned by the eigenstates of a spin-half $\sigma^z$ operator, as a function of the range of interactions and their strength. As we shall see, the structure of the density of states of our model makes the analysis of the gap ratio (and its mean), a typical measure in MBL studies \cite{Alet18,Sierant24}, not very informative. That is due to an approximate Hilbert space fragmentation in the interaction-dominated regime \cite{Sala20, Khemani20, moudgalya2022quantum}.

Because of the stronger interactions, the hopping or spin-flip is less probable, and Hamiltonian eigenstates appear more similar to the spin-aligned states or the eigenstates of $\sigma^z$ operators. The alignment is expected to become more robust and locally conserved as we move from left to right of the chain with linearly growing interaction strength. This phenomenon is similar to the Hilbert space fragmentation leading to the presence of almost conserved local operators \cite{Sala20, Khemani20, moudgalya2022quantum}. Therefore,  stronger interactions reduce the Hilbert space occupation, i.e., the localization volume of individual eigenstates---as observed in our work and as reported recently \cite{wang2024eigenstate}. This happens irrespective of the range of interactions. On the other hand, a long range of interactions enhances long-range correlations between spins.  
 Therefore, one always expects to see  an increase in the localization volume with an increase of interaction range. Surprisingly, however, we observe that the localization volume reduces upon increasing the range, and after an intermediate range it starts to grow again. Such a nonmonotonic behavior was not reported before in any clean (disorder-free) lattice model. We explain this phenomenon by moving from a spin picture to a particle occupation picture. We show that there is an interplay between the fragmentation induced by two-body interactions and an effective tilt potential induced by longer-ranged interaction.

The paper is organized as follows. We introduce our model and the measures studied in Sec.~\ref{sec:model}. Section \ref{sec:results_clean} is dedicated to our numerical full diagonalizations, the extraction of relevant results for clean systems with various ranges of interactions, and an explanation of the results. Section \ref{sec:qdisorder} is dedicated to the effect of quasi-periodic additive disorder on clean systems. We conclude in Sec.~\ref{sec:conclu}. Appendix \ref{app:schw} describes the connection of our clean Hamiltonian with the lattice gauge theory. Appendix \ref{appsec:fragmentation} discusses the Hilbert space fragmentation for perturbative hopping and outlines the Schrieffer-Wolff transformation to derive effective Hamiltonians.   

\section{Model}
\label{sec:model}

Here we define the  spin-half Hamiltonian on a 1D finite chain with various ranges of tilted ZZ interaction,

\begin{align}
 \mh_R = -t\sum_{n = 0}^{L-2} (\sigma^+_n \sigma^-_{n+1} + \sigma^-_n \sigma^+_{n+1}) \notag\\
 + U \sum_{l = 1}^{R} 
 \sum_{n = 0}^{L - (l+1)} (n+l) \sigma^z_n \sigma^z_{n+l}\;,
 \label{eq:ham_clean}
\end{align}
where the range $R$ can take any value from the integer set $\{1, 2, 3, \cdots, L-2, L-1\}$. We use the notation $\sigma^+_n$ = {\footnotesize$\left(\begin{array}{cc} 0 & 1\\ 0 & 0                                                                                                                                                                                                                                                                                                                                            \end{array} \right)$} = $(\sigma^-_n)^\dagger$, $\sigma^z_n = \frac{1}{2}{\footnotesize\left(\begin{array}{cc} 1 & 0\\ 0 & -1                                                                                                                                                                                                                                                                                                                                                  \end{array} \right)}$ for all lattice site $n$. $U$ is the interaction strength parameter, while $t$ is the hopping amplitude set to unity from now on.

Note that the interaction strength in the model \eqref{eq:ham_clean} depends not only on the relative distance  $l= (n+l) - n$ between spins but also depends on the spins' location. The nearest neighbor case of our model, $R=1$, is similar to the recently studied Hamiltonian in Ref.~\cite{wang2024eigenstate}. On the other hand, for ${R} = L-1$ our model is similar to the lattice Schwinger model of staggered spinless fermions after integrating out the electromagnetic gauge degrees of freedom---see Appendix \ref{app:schw} for details. For arbitrary $R$ the model may be, in principle, realized using cold ion quantum simulator architecture \cite{blatt2012}. For that reason, as well as having in mind possible Schwinger model realizations in quantum simulators \cite{Banuls20}, we discuss mostly systems with a moderate size $L$.

Still it is also interesting to consider the thermodynamic limit of the model. While for short-range interactions ($R$ small) such a limit seems straightforward, it is by no means so when $R$ approaches $L$. In this limit, the Hamiltonian \eqref{eq:ham_clean} seems to be superextensive. To remedy this problem, one can modify the interaction
strength as $U=\tilde{U}/R$ for $R$ close to $L$ and consider the properties of the Hamiltonian  at fixed $\tilde{U}$. In such a Kac-like rescaling, a growing range of interaction is now balanced by an effective reduction of interaction strength \cite{kac1963van}. This case resembles the extensivity problem for the model of cold atoms coupled by the cavity mode leading to an effective all-to-all interaction \cite{Chanda21self,Botzung21,Chanda22cav}.
Interestingly, in the case of the lattice gauge theory, the long-range interactions between matter particles appear after integrating out the electric flux using Gauss's law, which is local. Therefore, Kac's rescaling does not naturally arise in such cases, and we will not analyze it here.

The Hamiltonian [Eq.~\eqref{eq:ham_clean}] possesses two obvious symmetries:  It commutes with and hence dynamically conserves associated eigenvalues of (i) total magnetization operator $\sum_{n = 0}^{L-1} \sigma^z_n$ and (ii) the product of all spin-flip operators $\prod_{n = 0}^{L-1} \sigma^x_n$. 
We consider an even number of lattice sites and only a zero magnetization sector $\sum_{n = 0}^{L-1} \sigma^z_n = 0$ such that the total number of spin-up equals the total number of spin-down. This corresponds to the maximum Hilbert space dimension compared to other magnetization sectors, and it is similar to the half-filled spinless fermions. The zero magnetization sector can be divided into two mutually orthogonal vector spaces concerning the $\mathbb{Z}_2$ symmetry induced by the product of spin-flip operators: they consist of ``linear superpositions'' of $\prod_{n = 0}^{L-1} \sigma^z_n$ eigenstates which are either symmetric, having eigenvalue $+1$, or antisymmetric, having eigenvalue $-1$, under the action of  $\prod_{n = 0}^{L-1} \sigma^x_n$. The total number of basis states $\ket{s_j}$ in the symmetric sector is equal to that in the antisymmetric sector. Therefore, the Hilbert space dimension of one of the sectors is 
\begin{align}
 \md = \frac{L!}{2 [(L/2)!]^2}\;.\label{eq:hil_dim_clean}
\end{align}
We consider the symmetric sector only while analyzing the clean system. For the system size $L = 18$ and $16$, $\md$ = 24310 and 6435, respectively. We numerically diagonalize the full Hamiltonian in this sector for different $R$, $U$, and for different chain lengths $L$ with open boundary conditions. We concentrate on the second participation entropy \cite{mace2019multifractal} for the eigenvector $\ket{\psi_E}$ at each eigen-energy $E$,
\begin{align}
 \mathcal{S}(E) = -\ln\left(\sum_{j = 1}^\md \left|\langle s_j| \psi_E\rangle \right|^4 \right), 
\end{align}
calculated in the $\{\ket{s_j}\}$ basis defined above. For the mutually unbiased basis represented by  the set $\{\ket{s_j} : 1\leq j \leq \md\}$ and the set of eigenvectors $\{\ket{\psi_E}\}$ represented in this basis, each basis vector contributes with the same weight to $\ket{\psi_E}$ in a fully ergodic system, implying $\mathcal{S}(E)$ = $\ln\md$ \cite{karol}. In the opposite case with singly occupied  $\ket{\psi_E}$ = $\ket{s_j}$ state,  $\mathcal{S}(E) = 0$.  This implies $\mathcal{S}(E)/\ln\md \in [0,1]$. Therefore, the $\mathcal{S}(E)$ captures the localization volume in symmetric spin-space with zero magnetization.

\section{Numerical results}\label{sec:results_clean}

To take care of the energy dependence of our problem, we follow the by now standard procedure \cite{luitz2015many} and scale the energy eigenvalues, $E$, as 
\begin{align}
 \varepsilon = (E - E_{\min})/(E_{\max} - E_{\min}) \in [0,1]\;.
 \label{eq:scaled_e}
\end{align}

\begin{figure}[h]
\subfigure[]{\includegraphics[width = 0.43\textwidth]{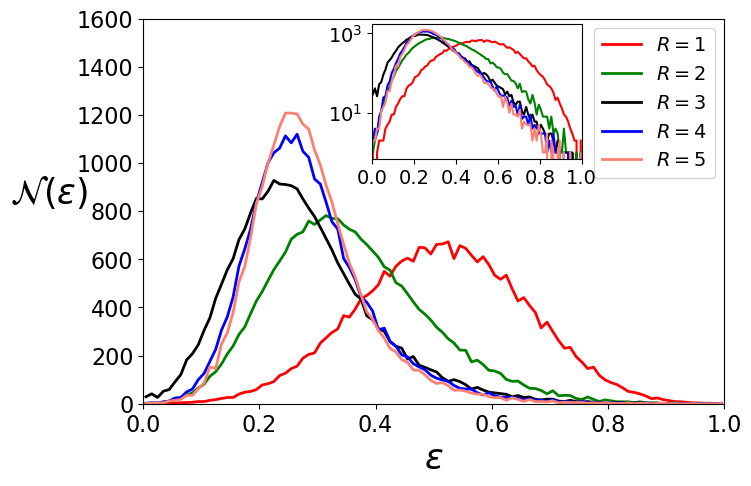}} 
\subfigure[]{\includegraphics[width = 0.43\textwidth]{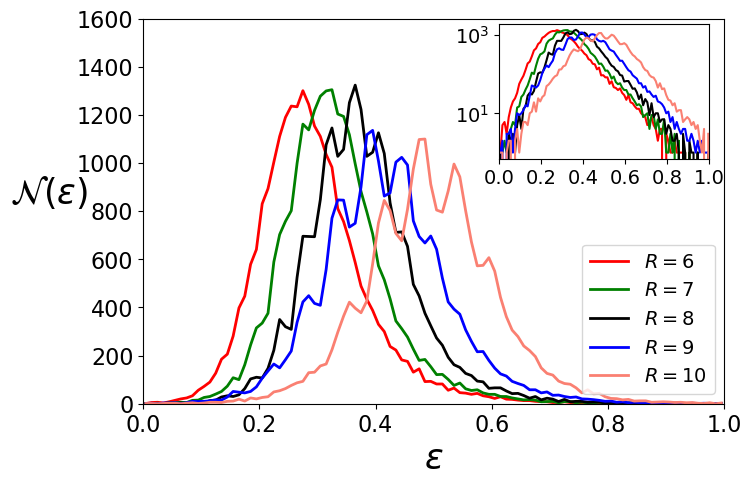}} 
\subfigure[]{\includegraphics[width = 0.43\textwidth]{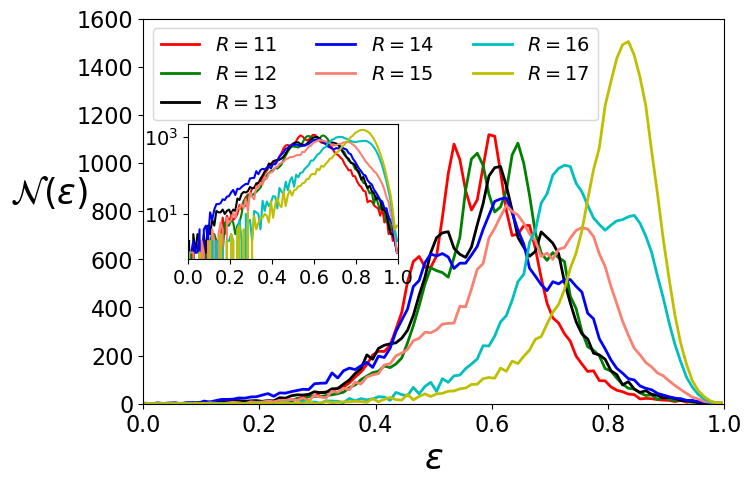}} 
\caption{Envelope plots of the histograms of the number of states $\mathcal{N}(\varepsilon)$ as a function of $\varepsilon$ [Eq.~\eqref{eq:scaled_e}] for different $R$ as indicated in the panels, while $U = 1.2$ and $L = 18$ for all subplots. 100 bins are used during histogram analysis for all cases. The insets are the same plots as in the main panels but on a log-linear scale.}
\label{fig:number_of_state}
\end{figure}

Figure \ref{fig:number_of_state} shows the number of states at different energies with the variation of range for $L = 18$ and $U = 1.2$. Two observations follow. First, with the increase of the range of interactions, the maximum of the density shifts first to smaller $\varepsilon$ values, then it comes back near to the center again, and, for still larger $R$, it moves towards $\varepsilon = 1$. Second, for intermediate $R$, distinct peaks appear in the $\mathcal{N}$ profile, which indicates that the interaction term in the Hamiltonian becomes dominant. In the limit of $t/U \rightarrow 0$, the spectrum splits trivially into exponentially many subsets of degenerate states corresponding to discrete values of the interaction in the spin bases---see Appendix \ref{appsec:fragmentation} for details.
In the presence of tunneling the degeneracy is lifted, leading still, however, to a multipeaked density of states (e.g.~Fig.~\ref{fig:number_of_state}). This behavior resembles to some extend an approximate Hilbert space fragmentation effect \cite{Sala20, Khemani20, moudgalya2022quantum} similar to that observed for strongly interacting dipoles in optical lattices \cite{Li21}. In the latter case, the Hilbert space fragmentation was due
to taking nearest-neighbor and next-nearest-neighbor interactions into account.
For larger ${R}$, the peak heights as well as the number of peaks in a distribution reduce with an increase of ${R}$.

\begin{figure}[h]
\includegraphics[width = 0.45\textwidth]{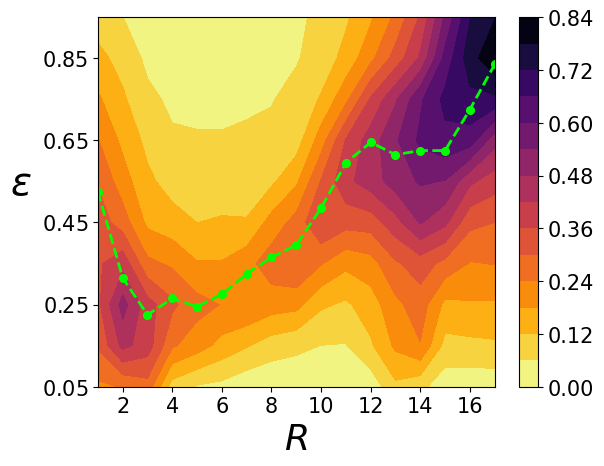}
\caption{Participation entropy $\mathcal{S}/\ln \md$ averaged over the energies within each bin in energy histogram (Fig.~\ref{fig:number_of_state}) as a function of scaled energy and range. $U = 1.2$, $L = 18$.  The dashed line connects the circular-marked data points, which signify the energy values $\varepsilon$ for which the $\mathcal{N}$ take maximum values in Fig.~\ref{fig:number_of_state}.} \label{fig:SE18}
\end{figure}
\begin{figure}[h]
\subfigure[]{\includegraphics[width = 0.45\textwidth]{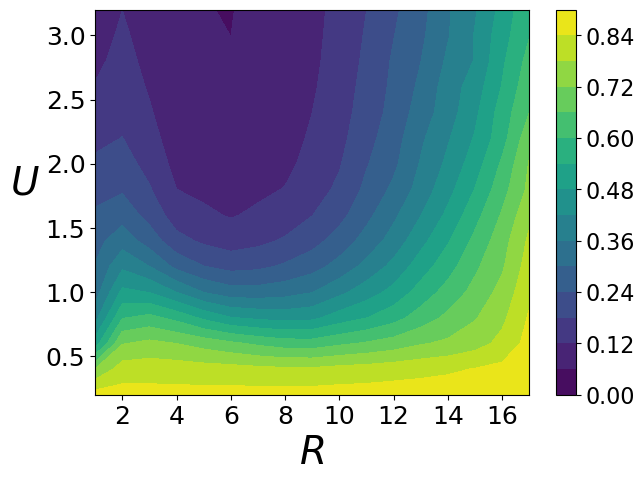}}
\subfigure[]{\includegraphics[width = 0.45\textwidth]{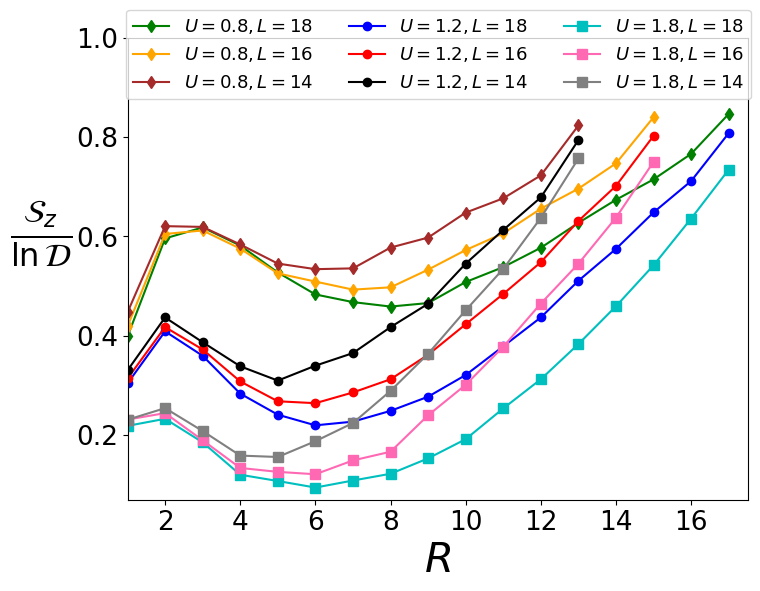}}
\caption{(a) Participation entropy $\mathcal{S}_Z/\ln \md$, i.e., $\mathcal{S}/\ln\md$ averaged over the set $Z$ [Eq.~\eqref{eq:def_90max_set}], as a function of $U$ and $R$. $L = 18$. (b) $\mathcal{S}_Z/\ln \md$ for different system sizes $L = 14~(\md = 1716), 16~(\md = 6435),
18~(\md = 24310)$, and different $U$.} \label{fig:Sm18}
\end{figure}

To analyze the energy-resolved participation entropy, we consider 10 bins in $\varepsilon$. 
We average the entropy over all eigenvalues within each bin. Figure~\ref{fig:SE18} shows that maximal $\mathcal{S}$ for a given $R$ appears, as might be anticipated, near the maximum peak in the density in Fig.~\ref{fig:number_of_state} for all $R$. Interestingly, however, for almost all energies at some intermediate values of ${R}$, the localization volume in spin Hilbert space, i.e., the participation entropy, is lower as compared to the values for shorter and longer ranges. The energy value $\varepsilon$ at which maximum localization volume occurs increases with the increase of the range for longer $R$, while the trend is inverted around $R = 3$. For the systems sizes $L = 14, 16, 18$ we found that the critical value $R$ for which the trend is inverted remains almost the same.

Next, we focus on the peak density of states by considering the following set of eigenvalues:
\begin{align}
 Z = \{\varepsilon: \ln[\mathcal{N}(\varepsilon)] \geq 0.9 \max\{\ln[\mathcal{N}(\varepsilon)]\} \}\;.
 \label{eq:def_90max_set}
\end{align}
The natural logarithm is taken to smooth the distributions---compare the insets and main plots in Fig.~\ref{fig:number_of_state} for details. The participation entropy averaged over all eigenstates that belong to  $Z$, labeled as $\mathcal{S}_Z$, is plotted as a function of $U$ and $R$ in Fig.~\ref{fig:Sm18}. 
The nonmonotonic or anomalous behavior of the localization volume as a function of range for almost all $U$ and all system sizes is visible. As shown in more detail in Fig.~\ref{fig:Sm18}{\color{red}(b)} at a fixed interaction strength $U$, a minimum of $\mathcal{S}_Z$ occurs at some intermediate range ${R} = {R}_c$, and the ${R}_c$ shifts to larger values with an increase of system size $L$. On the other hand, for a fixed $L$, an increase of $U$ shifts the ${R}_c$ to smaller values.  The thermodynamical behavior of $R_c$ will be discussed near the end of Sec.~\ref{subsec:discussion}.

\subsection{Localization transition analysis}

We now compare the participation entropies for different system sizes in units of $\ln \md$ to extract possible system-size-independent crossing points similar to the analysis done in \cite{mace2019multifractal}.

\begin{figure}[h]
\subfigure[]{\includegraphics[width = 0.235\textwidth]{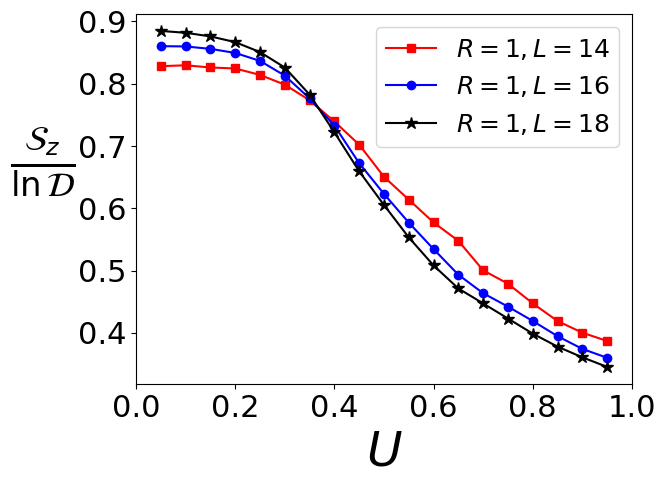}}
\subfigure[]{\includegraphics[width = 0.235\textwidth]{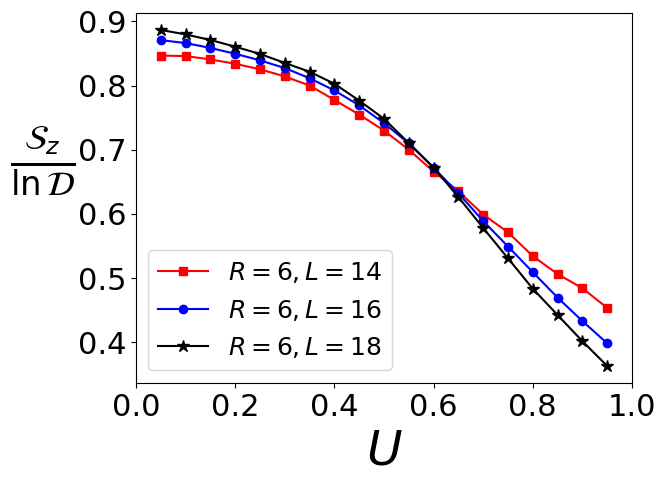}}
\subfigure[]{\includegraphics[width = 0.235\textwidth]{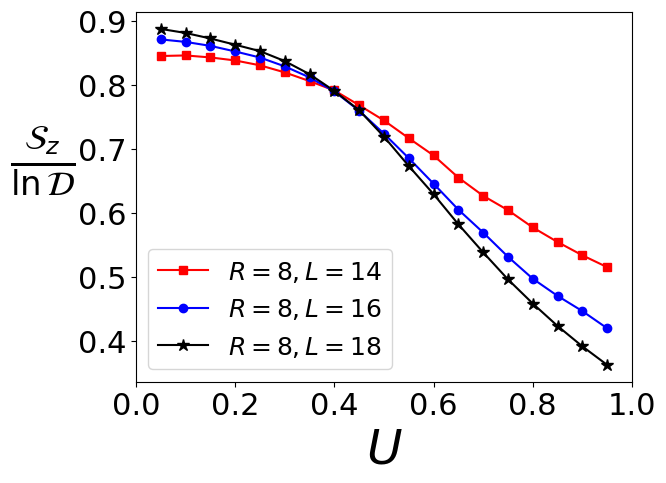}}
\subfigure[]{\includegraphics[width = 0.235\textwidth]{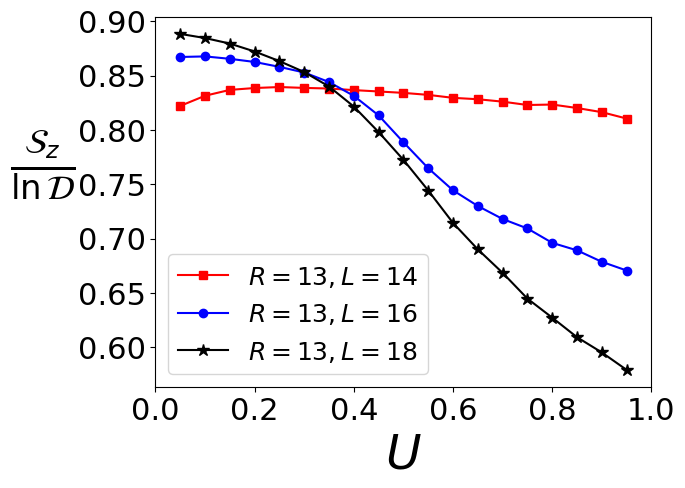}}\\
\subfigure[]{\includegraphics[width = 0.235\textwidth]{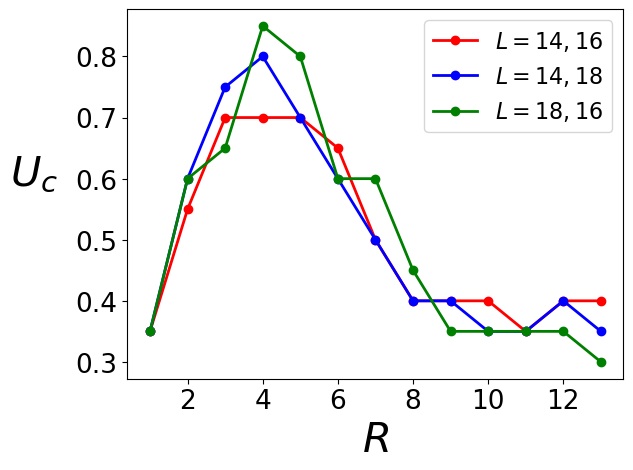}}
\subfigure[]{\includegraphics[width = 0.235\textwidth]{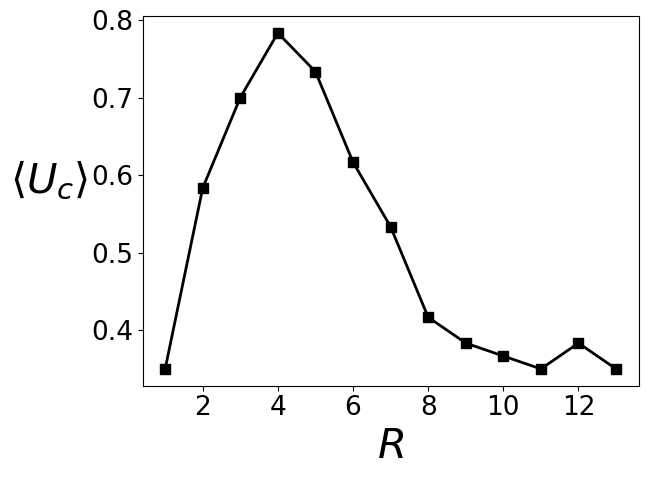}}
\caption{(a)-(d) $\mathcal{S}_Z/ \ln \md$ as a function of interaction parameter $U \geq 0.05$ for various $L$ and $R$. The crossing point at each subplot is associated with the ergodic-to-nonergodic transitions. (e) The transition interaction parameter $U_c$ as function of range extracted from different $L$ combinations. (f) Averaged $U_c$ over all combinations in (e).}
\label{fig:trans_sm}
\end{figure}

We plot $\mathcal{S}_Z/\ln\md$ as a function of interaction parameter $U$ for different ranges and system sizes in Figs.~\ref{fig:trans_sm}{\color{red}(a)}--\ref{fig:trans_sm}{\color{red}(d)}: only four $R$ cases are shown for convenience. The crossing point for different $L$ appears approximately at the same point $U = U_c$ for any fixed $R$. A closer inspection reveals that it also shows a nonmonotonic behavior with $R$. For small  $R$ the crossing point happens for smaller values of $U$, it shifts to larger $U$ at intermediate ranges, and for longer ranges it again moves to smaller $U$ values. The crossing points are associated with the ergodic-to-nonergodic transitions \cite{mace2019multifractal}. We conjecture that the crossing point $U_c \to 0$ when $R \to \infty$, which is possible at $L \to \infty$, and it signifies the appearance of nonergodic behavior for arbitrarily small $U$ in massless clean Schwinger-like models. To calculate the value of $U_c$ for each $R \leq 13$, we consider three crossing points of three pairs of curves: (i) $\mathcal{S}_Z/ \ln \md$ for $L = 14$ and $L = 16$, (ii) $\mathcal{S}_Z/ \ln \md$ for $L = 14$ and $L = 18$, and (iii) $\mathcal{S}_Z/ \ln \md$ for $L = 16$ and $L = 18$. Numerically, the crossing point is defined as the data coordinate $(U,\mathcal{S}_Z/ \ln \md)$ where the absolute value of the difference of $\mathcal{S}_Z/ \ln \md$ for two different $L$ is the minimum. The result is depicted in Figs.~\ref{fig:trans_sm}{\color{red}(e)} and \ref{fig:trans_sm}{\color{red}(f)}, which clearly captures a nonmonotonic behavior with $R$. For stronger $U$ we have shown the participation entropy plots in  Appendix \ref{app:sz_plots}. 

The anomalous effect is also visible in the mean gap ratio statistics as described in Appendix~\ref{gapratio}.

\begin{figure}[h]
\includegraphics[width = 0.4\textwidth]{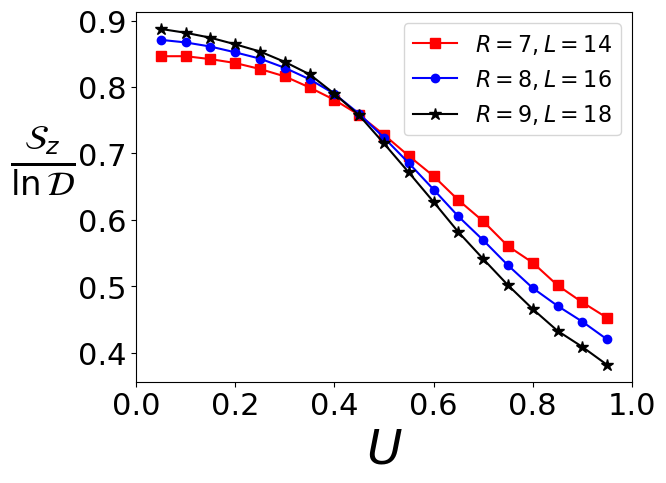}
\caption{{$\mathcal{S}_Z/\ln\md$ as a function of interaction parameter $U$ for different $L$ keeping the ratio $R/L = 1/2$}.}
 \label{fig:trans_sm_half}
 \end{figure}
We extract the interaction strength corresponding to the transition, $U_c = U_c(R)$, based on the comparison among different system sizes for fixed $R$, which is the standard practice for small $R$. Since the range $R$ may take any integer value up to $(L-1)$, another way to address the transition is by comparing $\mathcal{S}_Z/ \ln \md$ for different system sizes at fixed $R/L$ ratios. Because of the discrete nature of $R$, such an analysis requires data from different $L$ that are typically larger than accessible in exact diagonalization. Therefore,  the functional dependence $U_c = U_c(R/L)$ remains an open problem. We could present  such an analysis for the $R/L = 1/2$ case as shown in Fig.~\ref{fig:trans_sm_half}. It is qualitatively similar to Figs.~\ref{fig:trans_sm}{\color{red}(a)}--\ref{fig:trans_sm}{\color{red}(d)}.

\subsection{Discussion} \label{subsec:discussion} 
To explain the anomalous localization behavior, it is convenient to switch from the spin language to an equivalent spinless fermion (or hardcore boson) description. We define the particle number operator at site $n$ as $\hat{N}_n = \sigma^z_n + \frac{1}{2},$ relating the spin-up (spin-down) state to the presence (absence) of a particle. Only single occupancy at a lattice site is possible. Note that these spinless fermions are different from particles (or antiparticles) appearing in the staggered Schwinger picture discussed in Appendix \ref{app:schw}. The $\sigma_n^{\pm}$ operators are transferred into creation/annihilation operators for particles, and the first line in Eq.~\eqref{eq:ham_clean} gives the standard kinetic energy term (describing hopping). 
The interaction part becomes  
 \begin{align}
 \mh_\mathrm{in} =  &~ U \sum_{l = 1}^{R} 
  \sum_{n = 0}^{L - (l+1)} (n+l) \sigma^z_n \sigma^z_{n+l} \notag\\
  = &~ U \sum_{l = 1}^{R} 
  \sum_{n = 0}^{L - (l+1)} (n+l) \hat{N}_n \hat{N}_{n+l}  \notag\\&
  - \frac{U}{2} \sum_{l = 1}^{R} 
  \sum_{n = 0}^{L - (l+1)} (n+l) \big(\hat{N}_n + \hat{N}_{n+l}\big),\label{eq:inter_ham}
  \end{align}
  where we neglected the additional constant interaction energy term. The first term above is the two-body interaction operator. The second term strongly resembles the tilted potential present in recent studies of disorder-free many-body localization \cite{nieuwenburg2019bloch, schulz2019stark, Taylor20, yao2021nonergodic, Yao21}.  We use the term ``tilted potential'' referring to the single-particle potential term in the last line of Eq.~\eqref{eq:inter_ham}, and ``two-body interaction part'' to discuss the first term in Eq.~\eqref{eq:inter_ham}. The tilted interaction stands for the full ZZ-interactions term in Eq.~\eqref{eq:ham_clean}.

\textit{\textbf{Two-body interaction part}.} The two-body interactions  divide the available energy space into several clusters consisting of energy degenerate configurations. Different clusters are separated by energy $\sim U$. This leads to fragmented Hilbert space \cite{Yang2020hilbert, Sala20, Khemani20, moudgalya2022quantum}, which will be discussed in the following. We focus on the half-filled configurations, which are relevant for our work.
The important fact about the two-body interacting terms is that any particle at the right position $(n+l)$ has the same energy of interaction as any particle at the left site $(n+r)$ $<$ $(n+l)$ for different $r$
while the relative distance between them is $(l-r)$. In other words, the interaction operators $U (n+l) \hat{N}_n \hat{N}_{n+l}$ and $U (n+l) \hat{N}_{n+r} \hat{N}_{(n+r)+(l-r)}$ for a nonzero $r$ in Hamiltonian \eqref{eq:inter_ham} lead to the same interaction energy $U(n+l)$. Therefore, the absolute position of the left particle does not influence the two-body interaction term for a fixed position of the right particle at any relative distance. 
For example, if we represent the presence of a particle by ``$\footnotesize{\CIRCLE}$'' and the absence of a particle by ``$\footnotesize{\Circle}$'', the half-filled states $\ket{\footnotesize{\CIRCLE\Circle\Circle\Circle\CIRCLE\CIRCLE}}$ and $\ket{\footnotesize{\CIRCLE\Circle\Circle\CIRCLE\Circle\CIRCLE}}$ for $R = 2$ are degenerate having interaction energy = $5U$, and both of them have only one interacting pair. The presence of a particle in between the interacting pair (in the second example) changes energy configuration, i.e., $\ket{\footnotesize{\Circle\Circle\Circle\CIRCLE\CIRCLE\CIRCLE}}$ for $R = 2$ has three interacting pairs, and the total energy $= 14U$. Similarly, the two interacting pair configurations $\ket{\footnotesize{\CIRCLE\Circle\CIRCLE\Circle\CIRCLE\Circle}}$ and $\ket{\footnotesize{\Circle\CIRCLE\CIRCLE\Circle\CIRCLE\Circle}}$ for $R = 2$ are degenerate, having energy $= 6U$. Changing $R$ changes the distributions of the clusters.
\begin{figure}[h]
\subfigure[]{\includegraphics[width = 0.23\textwidth]{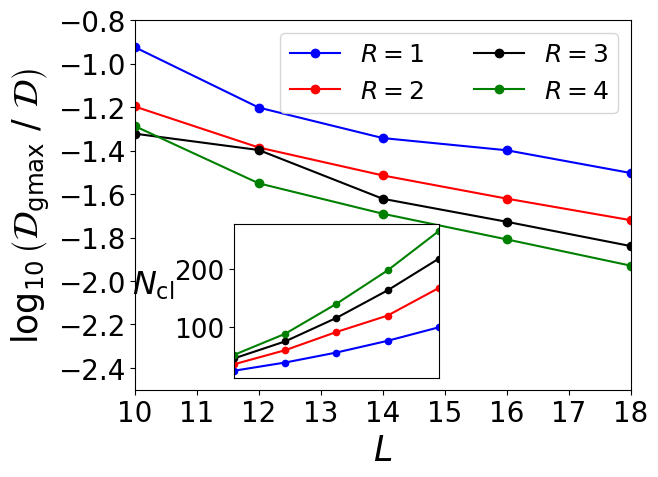}}
\subfigure[]{\includegraphics[width = 0.23\textwidth]{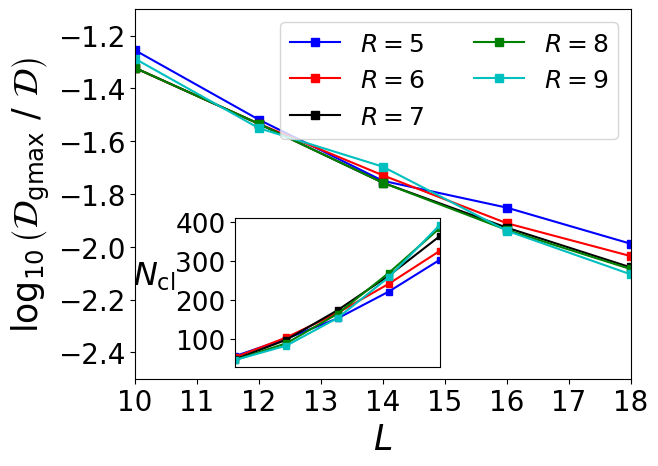}}
\caption{The ratio of dimension of the largest degenerate subspace = $\md_\mathrm{gmax}$ and Hilbert space dimension $\md$ as a function of system size $L$ and interaction range $R$ in the log-linear scale. The energy distributions are based on the two-body interacting Hamiltonian only (without hopping and tilted potential parts). The insets indicate the total number of clusters = $N_\mathrm{cl}$ as functions of $L$ for various $R$ with the same color lines as in the main plots.} 
\label{fig:dmax_frac}
\end{figure}

{For a typical range $R$, the number of such clusters grows exponentially with $L$, as identified in Fig.~\ref{fig:dmax_frac} for moderate system sizes.
The cluster with maximum dimension $ = \md_\mathrm{gmax}$ is exponentially smaller than the total dimension $\md$ of the Hilbert space of zero magnetization.
All the curves decay as $\approx 0.8^L$ with an increase in system size, indicating strong fragmentation. Because of the absence of spin-flip symmetry in the particle occupation language, we consider the symmetric and antisymmetric spaces together, i.e., here $\md$ is twice the dimension of the symmetric space [Eq.~\eqref{eq:hil_dim_clean}].}

\textit{\textbf{Tilted potential part.}} The tilted effective potential part in the particle picture for a general $R$ reads 
\begin{align}
V_R = &  - \frac{U}{2} \sum_{l = 1}^{R} 
  \sum_{n = 0}^{L - (l+1)} (n+l) \big(\hat{N}_n + \hat{N}_{n+l}\big)\notag\\
  = & - \frac{U}{2} \sum_{l = 1}^{R} 
  \Bigg[\sum_{n = 0}^{l-1} (n+l) \hat{N}_n + \sum_{n = L-l}^{L - 1} n \hat{N}_n\notag\\&
  ~~~~~~~~~ ~~~~ + \sum_{n = l}^{L - (l+1)} (2n+l) \hat{N}_n\Bigg]\notag\\
 = & - U \sum_{n = 0}^{L - 1} (R n) \hat{N}_n 
 + \frac{U}{2} \sum_{n = 0}^{R -1} n(R - n)\hat{N}_n \notag\\&
 + \frac{U}{4} \sum_{n = 0}^{R} \left[2 n L + n(n-1)\right]  \hat{N}_{L- R + n - 1}\;. 
 \label{eq:ws_pot}
\end{align}
For convenience, we denote $V_R/U$ as $\sum_{n = 0}^{L - 1} V(n) \hat{N}_n$. In the last two lines of the above expression \eqref{eq:ws_pot}, while changing the double index sums to single index sums, we have omitted the extra additive term = $-(UL/4)$$\sum_{l = 1}^R l$,  which comes from the total magnetization or particle number conservation: $\sum_{n=0}^{L-1} \hat{N}_n = L/2$ for even $L$. For detailed derivation, see Appendix \ref{app:expla_clean}. 

\begin{figure}[h]
\subfigure[]{\includegraphics[width = 0.23\textwidth]{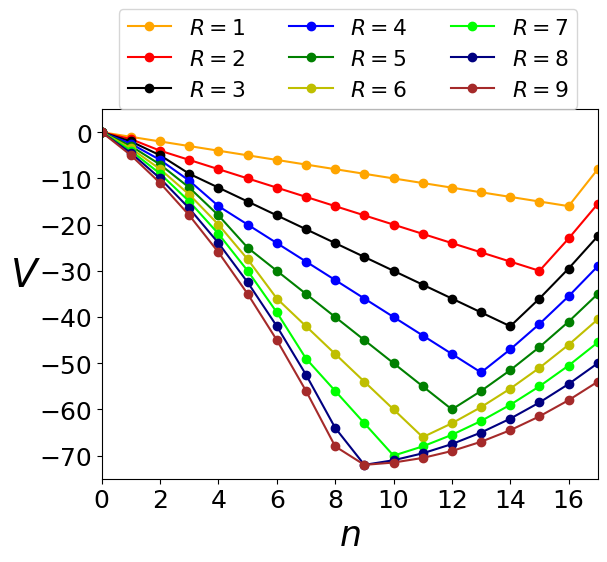}}
\subfigure[]{\includegraphics[width = 0.23\textwidth]{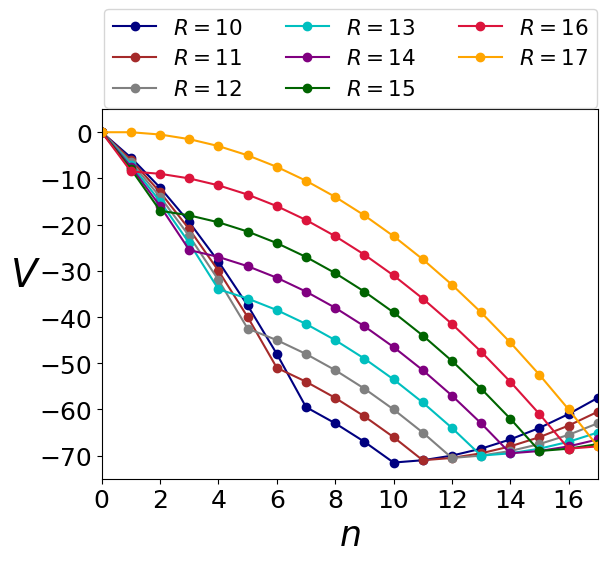}}
\caption{The tilted potential [Eq.~\eqref{eq:ws_pot}] as a function of lattice site $n$. The plots are for various ranges in units of $U$ for fixed system size $L = 18$. }
\label{fig:WS_energy}
\end{figure}

It is important to note that, after representing the single-particle potential in terms of single index sums, it contains both the linear and square terms in the site number. That complicates its analysis. The presence of the linearly tilted potential may induce another type of fragmentation in Hilbert space
due to an approximate conservation of the dipole moment, as extensively discussed
already \cite{Khemani20, Sala20}. Such a clean situation almost occurs for $R=1$; compare Fig.~\ref{fig:WS_energy}{\color{red}(a)}. However, the potential becomes quite different for a general $R$ forming first a triangular-like well with the minimum shifting close to the center of the lattice and changing its shape completely for longer ranges; cf.~Fig.~\ref{fig:WS_energy}. 
The $V_R$ effectively consists of three parts: the central region, and two sets associated with the sites near the two edges. Three different tilted parts become prominent for $R \geq 9$. Instead of linear tilt, they resemble two concave curves (left and central) and one convex curve (right). For longer  $R \geq 13$, the central region grows and the potential resembles a single concave quadratic curve.

\textit{\textbf{Simultaneous effect of two-body interaction and the tilted potential.}} 
\begin{figure}[h]
\subfigure[]{\includegraphics[width = 0.23\textwidth]{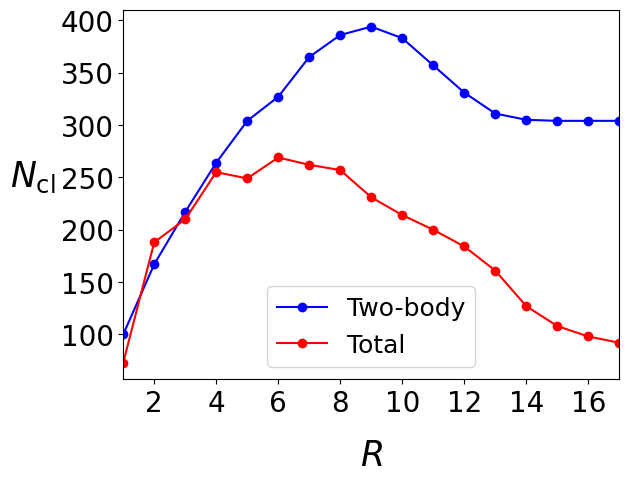}}
\subfigure[]{\includegraphics[width = 0.23\textwidth]{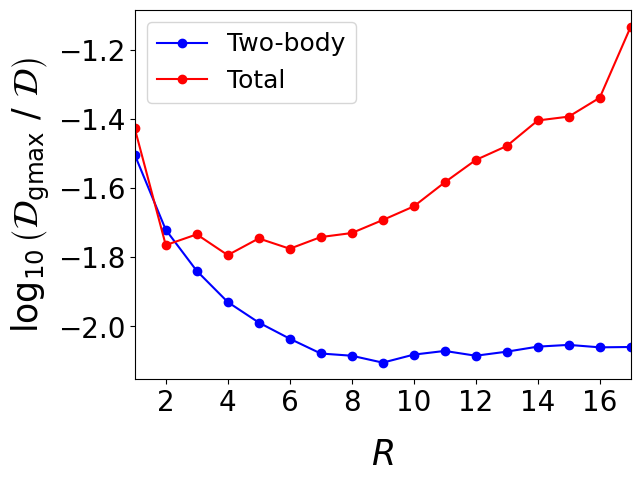}}
\caption{(a) Number of clusters or fragmented parts $N_\mathrm{cl}$ as a function of $R$ at $L = 18$. (b) The ratio of dimension of the largest degenerate space = $\md_\mathrm{gmax}$ and Hilbert space dimension $\md$ in log-linear scale as a function of $R$ at $L = 18$. Blue curves are in the absence of hopping and tilted potential, while red curves are in the absence of hopping only.}\label{fig:L18_frag_largeU}
\end{figure}
In the absence of hopping, Hilbert space fragmentation divides the whole system into several disconnected parts \cite{Sala20, Khemani20, moudgalya2022quantum}, while correlations exist among states inside each fragmented part upon turning on perturbative hopping $t/U$---see Appendix \ref{appsec:fragmentation} for a related discussion when $|t/U| \ll 1$. This fragmentation picture will be approximately valid when $U \gtrsim t$. In the presence of the tilted potential up to $R \leq 6$ when the positive gradient part of the potential ($\partial V/\partial n > 0$) grows with an increase in $R$ [Figs.~\ref{fig:WS_energy}{\color{red}(a)} and \ref{fig:WS_energy}{\color{red}(b)}], the number of fragmented subspaces $N_\mathrm{cl}$ grows with $R$; cf.~the red and blue curves in Fig.~\ref{fig:L18_frag_largeU}. 
Beyond $R > 8$, an increase of $R$ shrinks the positive gradient part in the tilted potential, which increases the number of states inside each fragmented part and the total number of fragmented parts $N_\mathrm{cl}$ reduces; cf.~the red and blue curves in Fig.~\ref{fig:L18_frag_largeU}. A similar effect is also visible from the growing subpeak volumes in the number of state distributions in Figs.~\ref{fig:number_of_state}{\color{red}(b)} and \ref{fig:number_of_state}{\color{red}(c)}. This qualitatively explains the decay of participation entropy $\mathcal{S}_Z$ with an increase in $R$ for $R < 7$ and the growing $\mathcal{S}_Z$ beyond $R \geq 7$ with an increase of $R$, e.g., Fig.~\ref{fig:Sm18} for a fixed $U$. Note that the blue curves in Fig.~\ref{fig:L18_frag_largeU} are without the restriction to the spin-flip symmetric subspace, as compared to the red curves. In the spin language without the hopping term $\equiv$ the red curves, the energy of any eigenstate from symmetric space will be the same as that of its antisymmetric partner eigenstate. Therefore, for the red curves, $N_\mathrm{cl}$ and the ratio $\md_\mathrm{gmax}/\md$ will be the same with and without the symmetric space restriction.       

 In Fig.~\ref{fig:Sm18} the critical $R = R_c$ increases with system size $L$ at fixed $U$, while at the same time it decreases with $U$ at fixed $L$. To make any proper conclusion, we require data from many larger system sizes (numerically inaccessible), because of the discreteness of the possible values of $R$. Nevertheless, we can try to estimate $R_c$ from the analytical expression Eq.~\eqref{eq:ws_pot} for the tilted potential, which is valid for arbitrary $R$ and $L$. We consider the perturbative $|t/U|$ regime. The highest number of fragmented subspaces appears around $R = 6$ in Fig.~\ref{fig:L18_frag_largeU}. Near the same order of $R \approx 8$, the triangular potential well formed by the tilted potential is prominent in Fig.~\ref{fig:WS_energy}, i.e., when the length of the lattice regions $L_p$, where the tilted potential has a positive gradient $\partial V/ \partial n > 0$, takes the maximum value. 
  \begin{figure}[h]
  \includegraphics[width = 0.4\textwidth]{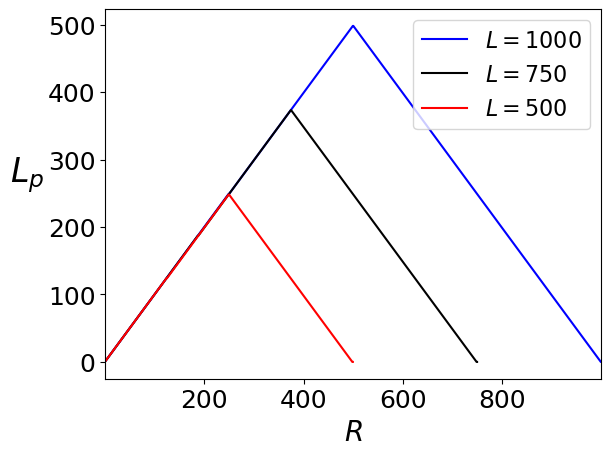}
  \caption{Length of the part of the lattice = $L_p$ where the tilted potential has positive gradient as a function of $R$ for different system sizes $L$.}
  \label{fig:L_p_vs_R}
 \end{figure}
  Figure \ref{fig:L_p_vs_R} shows the length of the part of the lattice equal to $L_p$, where the tilted potential has a positive gradient as a function of $R$ for different system sizes $L = 500, 750, 1000$. The $L - L_p$ part of the lattice contains a nonpositive gradient of the tilted potential. It is evident that the $L_p$ reaches maximum when $R/L \approx 0.5$. Therefore, we expect at the thermodynamic limit $R_c/L \lessapprox 0.5$. This is a rough estimation, and extra modifications will come from two-body interaction parts and the hopping, over the effect of triangular potential.

On the other hand, we observe an ergodic-to-nonergodic transition in Figs.~\ref{fig:trans_sm}{\color{red}(e)} and \ref{fig:trans_sm}{\color{red}(f)} where $U_c < 1$. In this case, the hopping strength $t = 1$ in the original Hamiltonian is relatively larger than the gaps between nearest fragmented clusters discussed above. This leads to mixing between different fragmented parts of the Hilbert space. Therefore, the previous explanation based on the fragmentation at the limit 
$|t/U| \ll 1$ will hardly work in this context. We keep the explanation for the nonmonotonic behavior of $U_c(R)$ as an open question.

\section{The presence of quasiperiodic disorder}
\label{sec:qdisorder}
The clean Hamiltonian is a diagonal matrix in its eigenbasis. In general, an additive disorder potential will not be diagonal in the same basis, and it will carry some off-diagonal parts that will act as tunneling between the eigenbasis states of the clean Hamiltonian. Therefore, disorder can enhance the localization volume in accessible Hilbert space compared to the clean localized system provided the latter is strongly localized. When the disorder strength becomes comparable to the gaps between nearest energy eigenvalues of the clean system, the localization volume reaches maximum. This is usually true when the disorder is weak compared to the
hopping and potentials of the clean Hamiltonian---cf.~a similar result in a flatband system \cite{goda2006inverse}. When the disorder becomes stronger, the eigenstate properties change completely, and one expects to see Anderson-localization-like phenomena. A similar feature can happen for our model.
To understand the anomalous localization behavior better and to check its robustness, we add quasiperiodic disorder on the original clean Hamiltonian [Eq.~\eqref{eq:ham_clean}]. The new Hamiltonian reads
\begin{align}
 \mh_\mathrm{dis}  = \mh_R +  \frac{W}{2}\sum_{n = 0}^{L-1} \cos(2\pi\beta n + \pi \varphi) \sigma^z_n\;,
 \label{eq:ham_dis}
\end{align}
where $\beta$ is the golden ratio = $(\sqrt{5} + 1)/2$, and the site-independent phase $\varphi$ is a random number picked from a uniform distribution $\in [0,1]$. This model is related to the truncated Schwinger model in the presence of disordered mass or chemical potential [Eq.~\eqref{eq:main_hamil_sch}]. Note that such on-site disorder breaks the spin-flip symmetry that was present in the clean Hamiltonian. Therefore, in the disorder case, instead of considering only the symmetric sector, we consider symmetric and antisymmetric blocks together, i.e., the dimension of the Hilbert space is now $\md = (L!)/[(L/2)!]^2$. The additive disorder term changes the structure of the potential function; cf.~$V_R$ [Eq.~\eqref{eq:ws_pot}].  Although it is impulsive to think that the disorder with homogeneous strength $W$ is negligible for the major right part of the chain because of the large tilted potential, it is not true in general.
In the presence of a linear potential, the eigenenergy levels are separated by a certain gap. Without interaction even in the presence of hopping, the eigenstates are localized around each lattice site. This is true irrespective of the location of the lattice site. The main point of interest comes when the disorder connects these different energy levels. In this case, the local potential from the linear part is not important as it can always be scaled down by subtracting an irrelevant constant potential; only the energy gap is important. For example, the eigenstate properties are the same for the two Hamiltonians $H = F \sum_{n = 0}^{L-1} n \hat{N}_n$ + hopping part and $H = F \sum_{n = 0}^{L-1} (n - L/2) \hat{N}_n$ + hopping part. The gap between nearest energy levels is $F$. The localization properties of the system start to change when the disorder strength $W \sim F$ \cite{GLUCK2002}.
Complexity arises through the presence of two-body interaction and the nonlinear part of the tilted potential [Eq.~\eqref{eq:ws_pot}]. 
\begin{figure}[h]
\subfigure[]{\includegraphics[width = 0.238\textwidth]{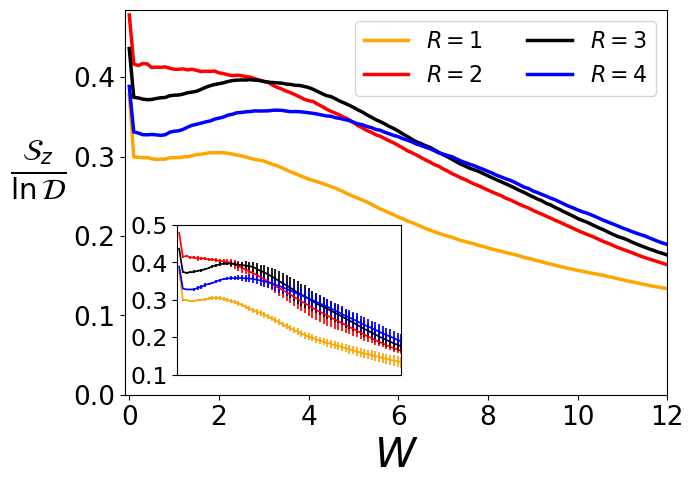}} 
\subfigure[]{\includegraphics[width = 0.238\textwidth]{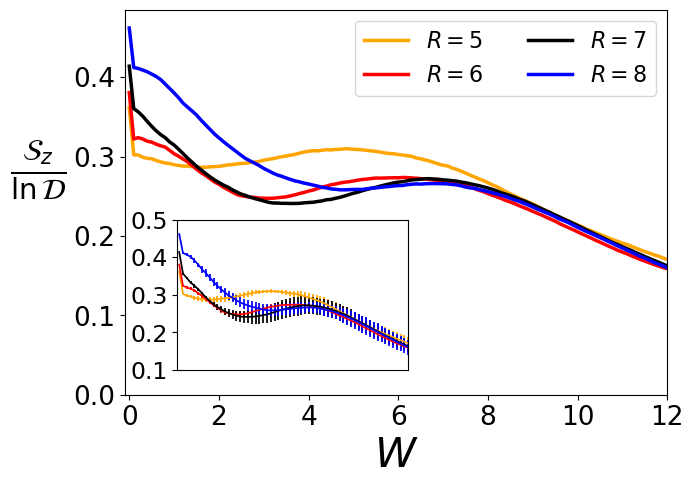}} 
\subfigure[]{\includegraphics[width = 0.238\textwidth]{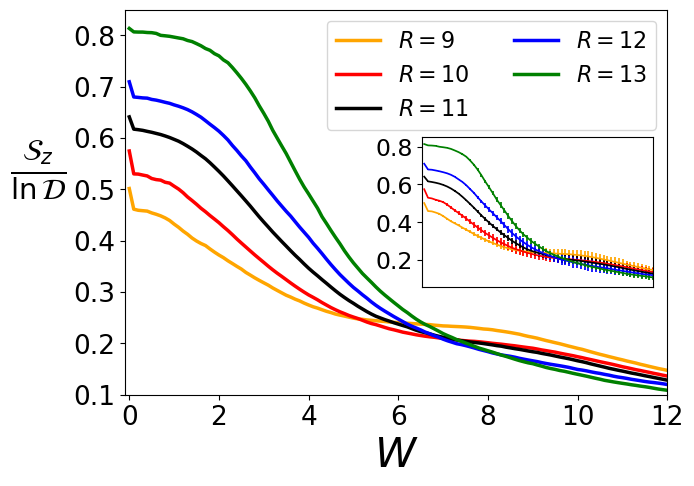}} 
\subfigure[]{\includegraphics[width = 0.23\textwidth]{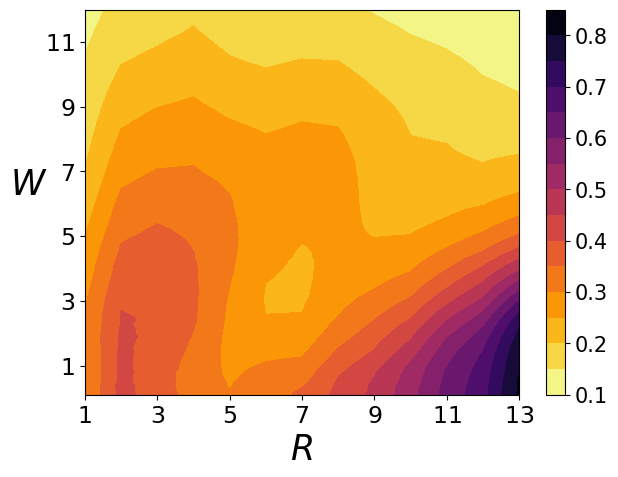}} 
\caption{Participation entropy $\mathcal{S}_Z$ which is averaged over the set $Z$ and over 150 random realizations of angle $\varphi$, as a function of $W$ and $R$. $L = 14$, $U = 1.2$, $\md = 3432$. Each inset is the same as the corresponding main plot but with error bars having vertical lengths two times the standard deviations calculated over the random realizations. Only two alternative data points are plotted in the insets for better visibility. In the last subfigure (d) we exclude the $W = 0$ cases.}
\label{fig:trans_sm_dis}
\end{figure}

For computational convenience, we choose $L = 14$ and $U = 1.2$ and show the participation entropies $\mathcal{S}$ as a function of disorder strength $W$ for variation of interaction ranges in Fig.~\ref{fig:trans_sm_dis}. The shown plots of $\mathcal{S}$ are averaged over the set $Z$ [Eq.~\eqref{eq:def_90max_set}] and scaled by the logarithm of the Hilbert space dimension $\ln \md$. 

Figure \ref{fig:trans_sm_dis} reveals that the same anomalous behavior of localization volume observed for the clean system is preserved at weaker disorder for all ranges. Introducing disorder with $W = 0.1$ on the clean system reduces the localization volume or $\mathcal{S}$ for all ranges of interaction. For stronger disorder, the system comes up with more exotic anomalous behaviors. For $R \leq 8$ [Figs.~\ref{fig:trans_sm_dis}{\color{red}(a)} and \ref{fig:trans_sm_dis}{\color{red}(b)}] we see an initial increase of localization volume with an increase of $W$, and a decrease of it upon a further increase of $W.$ In the approximate fragmentation picture, the disorder induces a transition between different parts of the fragmented Hilbert space, which results in growing correlations between them and the corresponding enhancement of the localization volume. The peaks occur when the disorder induces resonance. The increase of $\mathcal{S}$ is more prominent for the ranges $R \in \{1,3,4,\cdots,8\}$ when the fragmentation-induced localization is observed in Fig.~\ref{fig:Sm18}. The required $W$ increases for longer $R$ as the energy required to induce resonance between different fragmented sectors increases. For a very strong disorder, the system starts to show a monotonically decreasing localization volume, which is similar to the disorder-induced many-body localization.

\begin{figure}[h]
\subfigure[]{\includegraphics[width = 0.237\textwidth]{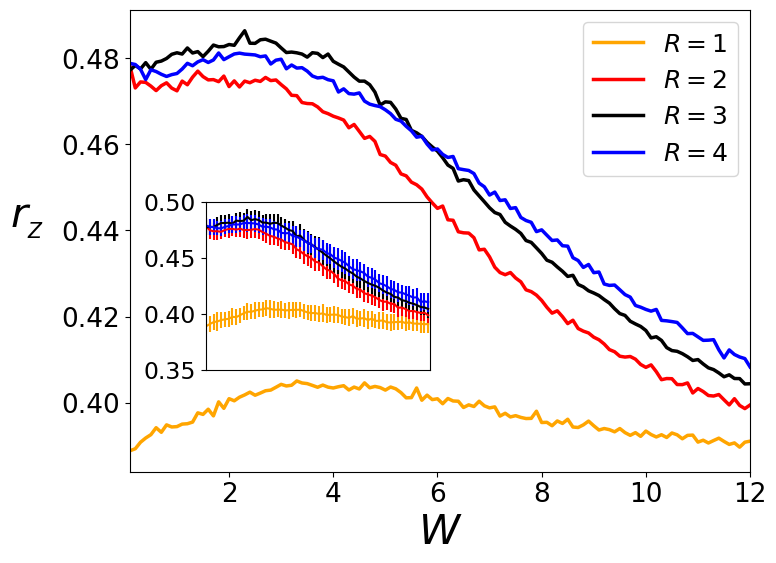}} 
\subfigure[]{\includegraphics[width = 0.237\textwidth]{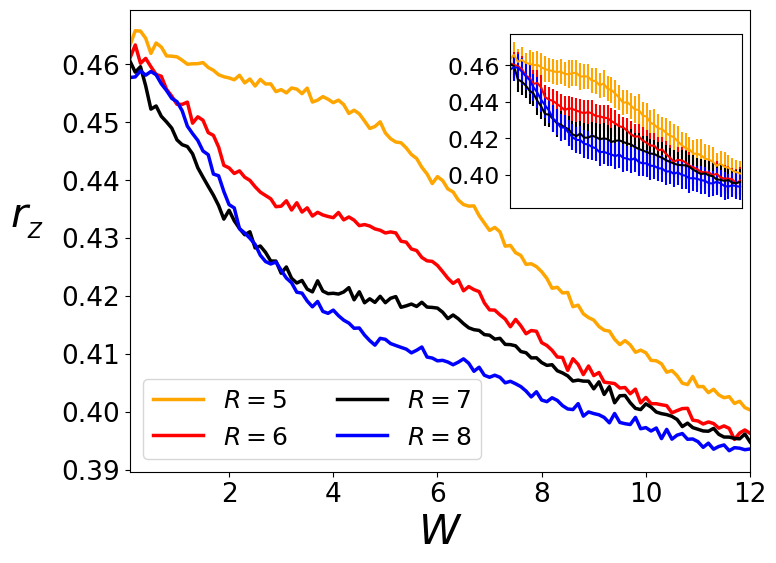}} \\
\subfigure[]{\includegraphics[width = 0.237\textwidth]{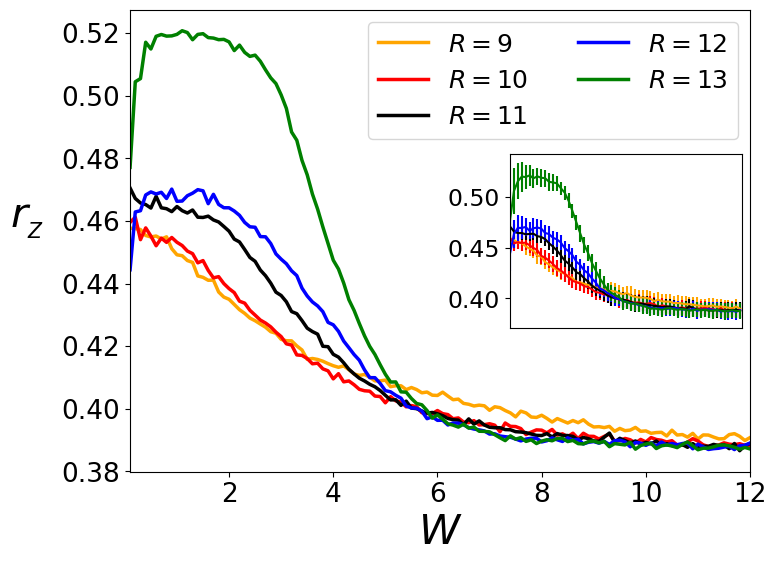}} 
\subfigure[]{\includegraphics[width = 0.237\textwidth]{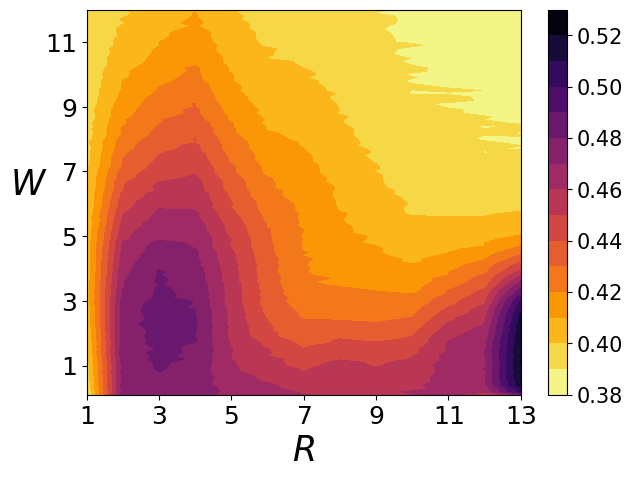}} 
\caption{Spectral gap ratios averaged over the set $Z$ and over 150 random realizations of angle $\varphi$, as a function of $W$ and $R$. $L = 14$, $U = 1.2$. Each inset is the same as the corresponding main plot but with error bars having vertical lengths two times the standard deviations calculated over the random realizations. Only two alternative data points are plotted in the insets for better visibility. We do not include $W = 0$ cases here.}
\label{fig:trans_rm_dis}
\end{figure}

The spectral gap analysis is shown in Fig.~\ref{fig:trans_rm_dis}. For the longest $R = (L-1)$ the system is in the ergodic regime for smaller $W$, while for larger $W$ it starts to enter into the nonergodic regime. For all other ranges at some intermediate value of $W$, we see either an increase of $r_Z$ or a slowdown in the transition from a maximum to minimum value. We relate this phenomenon with the disorder-induced local ergodicity or correlation between different fragmented sectors. For very large $W$, the system is near to a many-body localized regime.  

\section{Conclusion}
\label{sec:conclu}
We have discussed the localization properties of a 1D XXZ spin model with ZZ interaction strength dependent on the location of interacting spins for different ranges of interactions.
Our model resembles a lattice Schwinger model (quantum-electrodynamics with a zero time-component of the gauge potential) with a truncated range of interaction. The work shows that the localization properties do not change monotonically with the change of interaction range as one may naively expect. We explain this nonmonotonic behavior as an interplay between approximate Hilbert space fragmentation \cite{Sala20, Khemani20, moudgalya2022quantum} and an effective tilt potential induced by longer-ranged interaction. We were able to consider moderate system sizes so the exact scaling of the critical range of interaction $R_c$ for which the localization volume is minimum remains an open question. Based on the nature of the tilting potential, we roughly estimate a linear behavior: $R_c/L \lessapprox 0.5$, and an additional dependence from the hopping and two-body interaction terms may arise. The nonmonotonic effect remains unchanged in the presence of weaker disorder potentials. Our work contributes to the important question of many-body system properties under long-range interactions, which has an increasing physical relevance. It helps us to understand the nature of Coulomb-type interaction as well as the lattice gauge theory models in one dimension. The truncated or finite range of interactions may be associated with the presence of a charge screening or shielding effect.

In the future, we would like to extend to a higher-dimensional lattice where the nature of the interaction could change from linear to logarithmic or in different ways \cite{nandkishore2017manybody}. Introduction of sublattice structure and gauge potential flux can introduce different phenomena that need to be investigated.

It is known that the usual XXZ model is simulable by ultracold Rydberg atoms interacting by van der Waals forces \cite{ferracini2022realization}, where the atom states are mapped to spin-up and -down states. Tuning the van der Waals coefficients for different sets of two Rydberg atoms, we could map the inhomogeneous linearly growing ZZ long-range interactions while keeping the hopping term short-ranged [Eq.~\eqref{eq:ham_clean}]. One could use the recently implemented XXZ setups in cold $^7$Li atoms in optical lattices \cite{jepsen2020spin} where anisotropic ZZ interaction parameters can be set by locally controlling the applied magnetic field. It may also be possible to implement our model using a trapped ion-based digital quantum simulator \cite{nguyen2022digital}.

\begin{acknowledgments}
The authors acknowledge the financial support from the National Science Centre, Poland (Grant No.~2021/03/Y/ST2/00186) within the QuantERA II Programme, which has received funding from the European Union's Horizon 2020 research and innovation programme under Grant Agreement No.~101017733.  For the purpose of Open Access, the authors applied a CC-BY public copyright licence to any Author Accepted Manuscript (AAM) version arising from this submission. J.Z.~acknowledges also the partial support from the National Science Centre, Poland under Grant No.~2019/35/B/ST2/00034. A.M.~thanks Bitan De, Pedro Nicácio Falcão, and Adith Sai Aramthottil for useful discussions. We acknowledge Polish high-performance computing infrastructure PLGrid for awarding this project access to the LUMI supercomputer, owned by the EuroHPC Joint Undertaking, hosted by CSC (Finland) and the LUMI consortium through PLL/2023/04/016502.
\end{acknowledgments}

\appendix 

\section{Similarity of our Hamiltonian with the modified Lattice Schwinger Model}\label{app:schw}

The lattice Schwinger Hamiltonian for the staggered spinless fermions with the electromagnetic gauge potential reads \cite{kogut1975hamiltonian,banks1976strong} 
\begin{align}
 \mh_\mathrm{sch} = & - i\sum_{n=0}^{L-2} \left( \psi^\dagger_n e^{i \phi_{n,n+1}} \psi_{n+1} -   \psi^\dagger_{n+1} e^{-i \phi_{n,n+1}} \psi_{n} \right) \notag\\
 & + U \sum_{n=-1}^{L-1} \mathcal{E}^2_{n,n+1}  + \sum_{n=0}^{L-1} m_n (-1)^n \psi^\dagger_n \psi_{n}\;. \label{eq:main_hamil_sch}
\end{align} 
It controls the dynamics of spinless fermion and antifermion fields $\psi_n$ coupled to $U(1)$ gauge boson fields $\phi_{n,n+1}$, which act as vector
potential with the temporal gauge, and $\mathcal{E}_{n,n+1}$ is the corresponding electric field and conjugate momentum of $\phi_{n,n+1}$. In the lattice representation, fermion fields are located at lattice sites $n$ while the gauge fields are at the links between sites.
{ For the clean system we} consider the massless case: $m_n = 0$. The presence and absence of particles (or antiparticles) form two possibilities at each site $n$, i.e., Hilbert space of dimension 2. The dynamical fermionic charge operator is defined as 
\begin{align}
          &\mathcal{F}_n = \psi^\dagger_n \psi_{n} - \frac{1}{2}[1 - (-1)^n]\left(\begin{array}{cc}                                                                                                                                                                                                                            
           1 & 0\\ 0 & 1                                                                                                                                                                                                                                 \end{array}\right), ~~ \psi^\dagger_n \psi_{n} =  \left(\begin{array}{cc}                                                                                                                                                                                                                            
           1 & 0\\ 0 & 0                                                                                                                                                                                                                                 \end{array}\right) \notag\\
          &\implies \mathcal{F}_{n = \text{even}} = \psi^\dagger_n \psi_{n},~ 
          \mathcal{F}_{n = \text{odd}} = -\psi_n \psi^\dagger_{n}\;.
          \end{align} 
In this notation, the particle can only sit at an even site and it is annihilated by operator $\psi$, while the antiparticle can sit at the odd site only and it is annihilated by operator $\psi^\dagger$. In the presence of background static charges $q_n$, the discrete Gauss law reads

\begin{align}
\left( \mathcal{E}_{n,n+1} - \mathcal{E}_{n-1,n} - \mathcal{F}_n \right)\ket{\Psi} = q_n \ket{\Psi},
\end{align}
where $\ket{\Psi}$ are the only allowed physical states. If we set 
\begin{align}
 q_n = -\frac{(-1)^n}{2}, 
\end{align}
the Gauss law becomes 
 \begin{align}
\mathcal{E}_{n,n+1} -  \mathcal{E}_{n-1,n}  = \frac{1}{2}\left(\begin{array}{cc}                                                                                                                                                                                                                            
           1 & 0\\ 0 & -1                                                                                                                                                                                                                                 \end{array}\right) = \sigma^z_n\;. \label{appeq:gauss_law}
\end{align}

The Jordan-Wigner transformation \cite{hamer1997series} 
\begin{align}
\psi_n = \prod_{j=0}^{n-1} \left[2 i e^{-i \phi_{j,j+1}} \sigma^z_{j}\right] \sigma_n^{-},         
\end{align}
with Pauli spin matrices $\sigma^+_n$ = ${\footnotesize\left(\begin{array}{cc}                                                                                                                                                                                                                            
           0 & 1\\ 0 & 0                                                                                                                                                                                                                                 \end{array}\right)}$, $\sigma^-_n$ = ${\footnotesize\left(\begin{array}{cc}                                                                                                                                                                                                                            
           0 & 0\\ 1 & 0                                                                                                                                                                                                                                 \end{array}\right)}$, 
 and $\sigma^z_n$, causes the hopping part of the Hamiltonian to become 
 \begin{align}
 - \sum_{n = 0}^{L-2} \left(\sigma^+_n \sigma^-_{n+1} + \sigma^-_n \sigma^+_{n+1}\right), \label{appeq:sch_hopp}
 \end{align}
which is the XX part of the usual XXZ spin chain model. 

For our work, we modify the gauge field energy term of the lattice Schwinger model with link $g_n$ dependent terms  
\begin{align}
 U \sum_{n=-1}^{L-1} g_n \mathcal{E}^2_{n,n+1} 
\;. \label{appeq:electric_potential}\end{align}

The hopping part [Eq.~\eqref{appeq:sch_hopp}] and potential [Eq.~\eqref{appeq:electric_potential}] together form the modified lattice Schwinger Hamiltonian.

If we choose $\mathcal{E}_{-1,0} = 0$ from Gauss law [Eq.~\eqref{appeq:gauss_law}], we obtain
\begin{align}
 \mathcal{E}_{0,1} = \sigma^z_0 \implies \mathcal{E}_{1,2} = \sigma^z_1 + \sigma^z_0 \notag\\
 \implies \cdots 
 \implies \mathcal{E}_{n,n+1} = \sum_{j=0}^n \sigma^z_j \;.\label{appeq:electric_fluxes}
 \end{align}
 Therefore, the energy term becomes 
  \begin{align}
& \sum_{n=0}^{L-1} g_n \mathcal{E}^2_{n,n+1} = \sum_{n=0}^{L-1} g_n \left(\sum_{j=0}^n \sigma^z_n\right)^2 \notag\\
& = g_0 \left(\sigma^z_0\right)^2 + g_1 \left(\sigma^z_0 + \sigma^z_1\right)^2 \notag\\
 & + g_2 \left(\sigma^z_0 + \sigma^z_1 + \sigma^z_2\right)^2 
 + \cdots + g_{L-1} \left(\sum_{j=0}^{L-1} \sigma^z_j\right)^2 \;. \end{align}

We use the fact the $(\sigma^z_j)^2$ is proportional to the identity and hence can be treated as an additional constant and is unimportant, therefore up to these additional constants
  \begin{align}
 \sum_{n=0}^{L-1} g_n \mathcal{E}^2_{n,n+1}  = &~ 2   \sigma^z_0 \sigma^z_1 \sum_{j = 1}^{L-1} g_j 
 +  2  \left[\sigma^z_0 \sigma^z_2 +  \sigma^z_1 \sigma^z_2\right]\sum_{j = 2}^{L-1} g_j \notag\\&
 +  2  
  \left[\sigma^z_0 \sigma^z_3 + \sigma^z_1 \sigma^z_3 + \sigma^z_2 \sigma^z_3 \right] \sum_{j = 3}^{L-1} g_j \notag\\&
  + \cdots + 2 \left[\sum_{n = 0}^{L-2} \sigma^z_n \sigma^z_{L-1}\right] g_{L-1}\;.
\end{align}
To produce our model Hamiltonian [Eq.~\eqref{eq:ham_clean}] with the longest range of ZZ interaction $R = L-1$, we need to set 
\begin{align}
 &\sum_{j = p}^{L-1} g_j = \frac{p}{2};~~p = 1,2, \cdots, L-1\notag\\
 &\Rightarrow g_1 = g_2 = \cdots = g_{L-2} = -\frac{1}{2},~g_{L-1} = \frac{L-1}{2}\;.
\end{align}
 This is the gauge energy term in a lattice Schwinger Hamiltonian in the presence of a defect at a boundary. In the case of a total zero magnetization sector, Eq.~\eqref{appeq:electric_fluxes} implies $\mathcal{E}_{L-1,L} = 0$. Therefore, the defect will not appear in the potential part \eqref{appeq:electric_potential}.
 For the shorter interaction range $R < (L-1)$, the Hamiltonian [Eq.~\eqref{eq:ham_clean}] represents a truncated version of the modified Schwinger Hamiltonian.

\section{Plots of $\mathcal{S}_Z$ for the extended $U$}\label{app:sz_plots}
\begin{figure}[H]
\subfigure[]{\includegraphics[width = 0.237\textwidth]{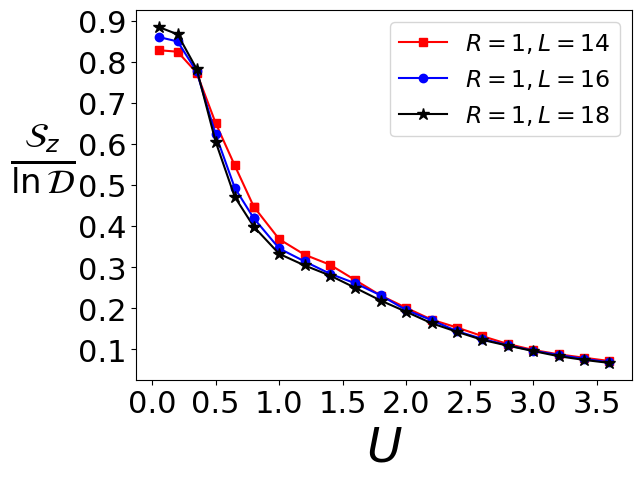}}
\subfigure[]{\includegraphics[width = 0.237\textwidth]{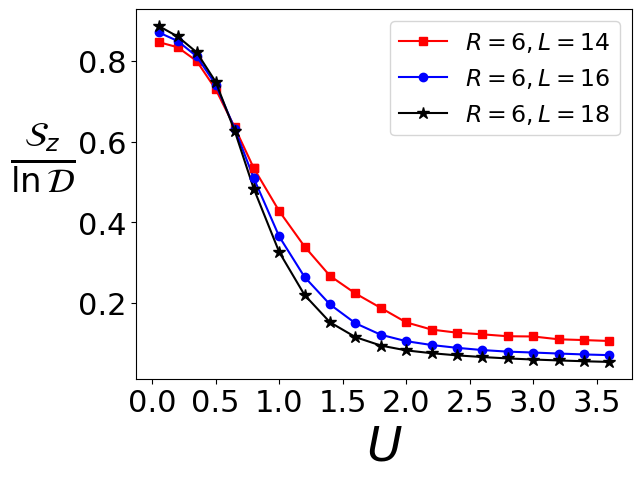}} \\
\subfigure[]{\includegraphics[width = 0.237\textwidth]{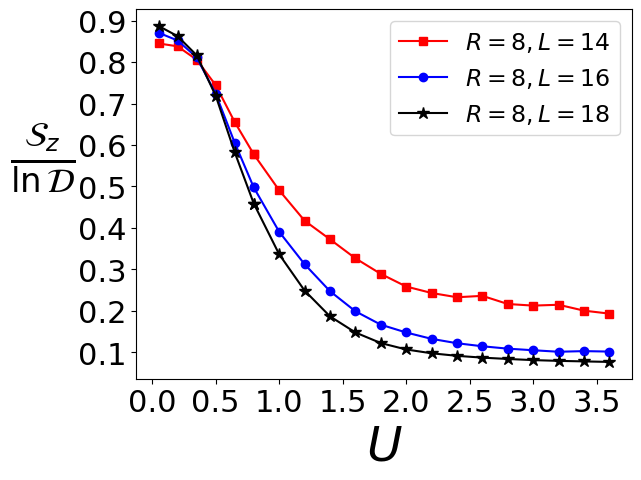}}
\subfigure[]{\includegraphics[width = 0.237\textwidth]{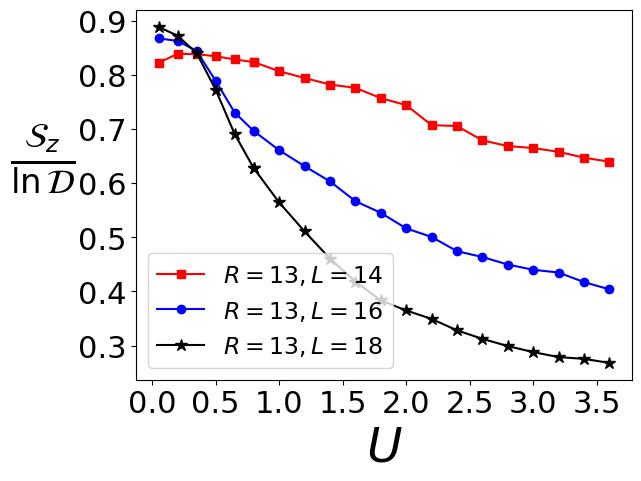}}
\caption{(a)-(d) $\mathcal{S}_Z/ \ln \md$ as a function of interaction parameter $U \geq 0.05$ for various $L$ and $R$ for the clean system.}
\label{fig:trans_sm_app}
\end{figure}

Figure \ref{fig:trans_sm_app} shows $\mathcal{S}_Z$ for the extended $U$.
\section{Mean gap ratio statistics}\label{gapratio}
\begin{figure}[H]
\subfigure[]{\includegraphics[width = 0.237\textwidth]{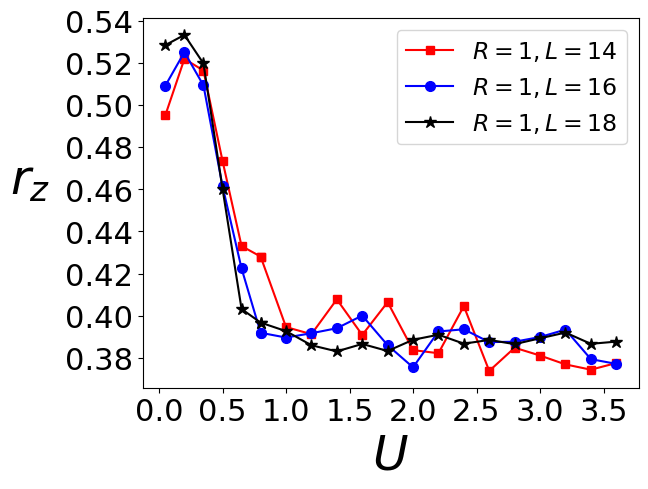}} 
\subfigure[]{\includegraphics[width = 0.237\textwidth]{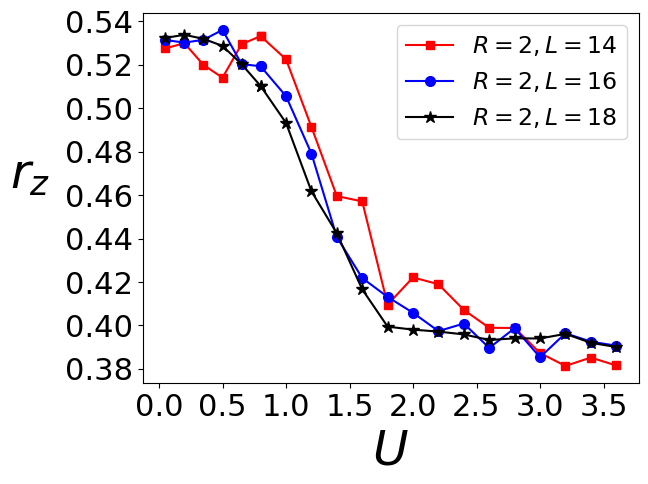}}\\
\subfigure[]{\includegraphics[width = 0.237\textwidth]{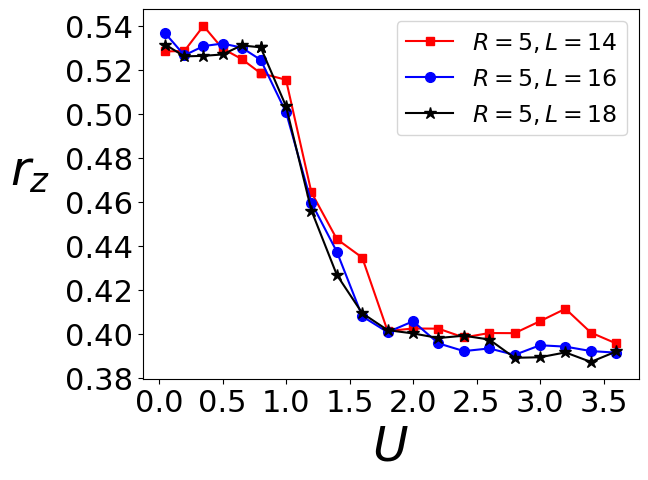}} 
\subfigure[]{\includegraphics[width = 0.237\textwidth]{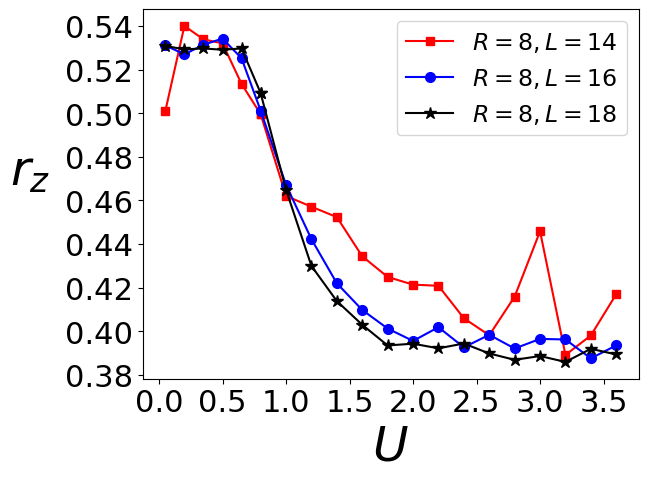}}\\
\subfigure[]{\includegraphics[width = 0.237\textwidth]{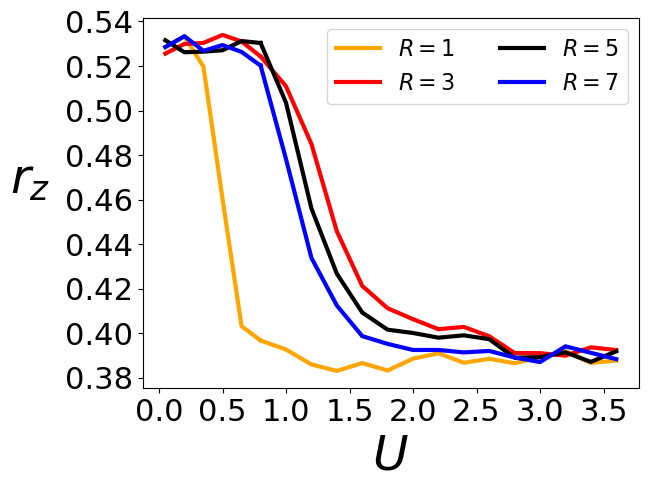}} 
\subfigure[]{\includegraphics[width = 0.237\textwidth]{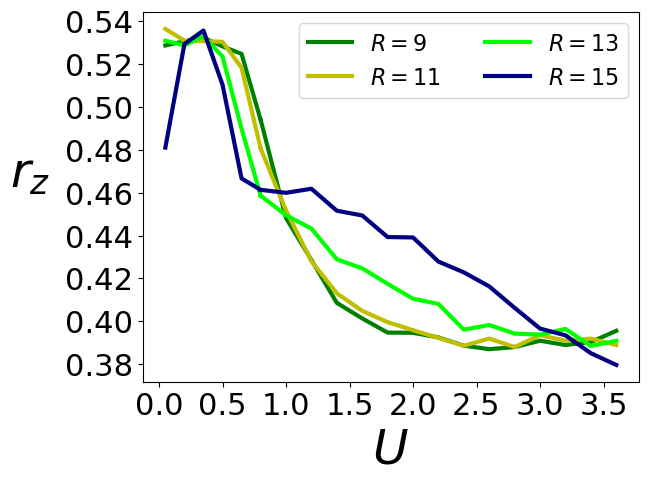}}
\caption{(a)-(d) Spectral gap ratio $r$ averaged over the set $Z$ as a function of $U$ for various $L$ and $R$. 100 bins are used for all cases. Subfigures (e),(f) are for $L = 18$ only.}
\label{fig:spectral_ratio_max}
\end{figure}

The mean gap ratio \cite{oganesyan2007localization,atas2013distribution,Sierant19,Sierant20} is a most useful and frequently used measure
to distinguish between ergodic and localized dynamics in many-body systems. We define the
dimensionless gap ratio of two subsequent spacings between eigenvalues $\varepsilon_{\alpha}$,
\begin{align}
 r_\alpha = \frac{\min\{\varepsilon_{\alpha+1} - \varepsilon_{\alpha}, \varepsilon_{\alpha} - \varepsilon_{\alpha-1}\}}{\max\{\varepsilon_{\alpha+1} - \varepsilon_{\alpha}, \varepsilon_{\alpha} - \varepsilon_{\alpha-1}\}},
\end{align}
with $\alpha$ being the index of the sorted eigenvalues in ascending order. The mean of $r_\alpha$ in the  eigenvalue interval studied
is $r_{GOE}\approx 0.53$ for the ergodic case and $r_{POI}\approx 0.386$ for the localized regular region \cite{atas2013distribution}.
Figure~\ref{fig:spectral_ratio_max} shows the spectral gap ratio $r$ averaged over all $\alpha$ for which $\varepsilon_\alpha, \varepsilon_{\alpha \pm 1}$ belong to the set $Z$ \eqref{eq:def_90max_set} as a function of $U$ for different ranges. We labeled it as $r_Z$. The crossing points $U_c$ in Figs.~\ref{fig:trans_sm}{\color{red}(e)} and \ref{fig:trans_sm}{\color{red}(f)} correspond to the $r_Z$ values corresponding to the entrance of the system from the ergodic regime to the nonergodic regime.  

 \section{DC field potential part [Eq.~\eqref{eq:ws_pot}] of the interacting Hamiltonian for a clean system}\label{app:expla_clean}
 
Here we derive the form of the tilted potential term for general $R$ after checking the expressions for a few special ranges. 

\subsubsection{$R = 1$}
We have 
{\small\begin{align}
 V_1 & = - \frac{U}{2} \sum_{n = 0}^{L - 2} (n+1) \big(\hat{N}_n + \hat{N}_{n+1}\big) \notag\\&
 = - \frac{U}{2} \hat{N}_0 - \frac{U}{2} (L-1) \hat{N}_{L-1} 
 - \frac{U}{2} \sum_{n = 1}^{L - 2} (2n+1) \hat{N}_n \notag\\&
 = \frac{UL}{2} \hat{N}_{L-1} 
 - \frac{U}{2} \sum_{n = 0}^{L - 1} 2n \hat{N}_n - \frac{UL}{4},
\end{align}}
where in the last term we used the total spin zero condition, which implies a half-filled lattice $\sum_{n} \hat{N}_n = L/2$. 
\subsubsection{$R = 2$}
We have 
{\small\begin{align}
V_2 & =  V_1 - \frac{U}{2} \sum_{n = 0}^{L - 3} (n+2) \big(\hat{N}_n + 
         \hat{N}_{n+2}\big)\notag\\&
 =  - \frac{U}{2}\big[2\hat{N}_0 + 3 \hat{N}_1 + (L-1) \hat{N}_{L-1} + (L-2) \hat{N}_{L-2}\big]\notag\\& ~~~
 - \frac{U}{2} \sum_{n = 2}^{L - 3} (2n+2) \hat{N}_n + V_1 \notag\\&
 = - \frac{UL}{2} - \frac{U}{2}\big[\hat{N}_1 + (L-3) \hat{N}_{L-1} + (L-4) \hat{N}_{L-2}\big]\notag\\& ~~~
 - \frac{U}{2} \sum_{n = 2}^{L - 3} 2n \hat{N}_n + V_1 \notag\\&
 = - \frac{U}{2}\big[3\hat{N}_1 + (2L-5) \hat{N}_{L-1} + (3L-8) \hat{N}_{L-2}\big] \notag\\& ~~~
 - \frac{U}{2} \sum_{n = 2}^{L - 3} 4n \hat{N}_n \notag\\&
 = \frac{U}{2}\big[\hat{N}_1 + (2L+1) \hat{N}_{L-1} + L \hat{N}_{L-2}\big]
 - \frac{U}{2} \sum_{n = 0}^{L - 1} 4n \hat{N}_n\;.
\end{align}}
In the penultimate line, we omitted the constant terms. 
 \subsubsection{$R = 3$}
We have 
{\small\begin{align}
 V_3 & = V_2  - \frac{U}{2} \sum_{n = 0}^{L - 4} (n+3) \big(\hat{N}_n + \hat{N}_{n+3}\big)\notag\\&
 =  - \frac{U}{2} \left[3 \hat{N}_0 + 4 \hat{N}_1 + 5 \hat{N}_2 
 + \sum_{p = 1}^3(L-p)\hat{N}_{L-p}\right]\notag\\& ~~~
 - \frac{U}{2} \sum_{n = 3}^{L - 4}(2n+3) \hat{N}_n + V_2 \notag\\&
 = - \frac{U}{2} \left[\hat{N}_1 + 2 \hat{N}_2 
 + \sum_{p = 1}^3(L-p-3)\hat{N}_{L-p}\right]\notag\\& ~~~
 - \frac{U}{2} \sum_{n = 3}^{L - 4} 2n \hat{N}_n - \frac{3UL}{4} + V_2 \notag\\&
 = - \frac{U}{2} \big[4\hat{N}_1 + 10 \hat{N}_2\big]
 - \frac{U}{2} \sum_{n = 3}^{L - 4} 6n \hat{N}_n \notag\\& ~~~
 -\frac{U}{2}\big[(3L-9)\hat{N}_{L-1} 
 + (4L-13)\hat{N}_{L-2} + (5L-18)\hat{N}_{L-3} \big] \notag\\&
 = - \frac{U}{2} \sum_{n = 0}^{L - 1} 6n \hat{N}_n  
 + \frac{U}{2}\big[2\hat{N}_1 + 2 \hat{N}_2
 + (3L + 3)\hat{N}_{L-1} \notag\\& ~~~
 + (2L+1) \hat{N}_{L-2} + L \hat{N}_{L-3} \big],
\end{align}}
where again we omitted the constant term coming from the total magnetization conservation. 

Therefore, for a general $R$ the resulting term reads
 {\normalsize\begin{align}
V_R = & - U R \sum_{n = 0}^{L - 1} n \hat{N}_n 
 + \frac{U}{2} \sum_{n = 0}^{R -1} n(R - n)\hat{N}_n \notag\\&
 + \frac{U}{4} \sum_{n = 0}^{R} \left[2nL + n(n-1)\right]  \hat{N}_{L- R + n - 1} \;.
\end{align}}

 \section{Hilbert space fragmentation and derivation outline for an effective Hamiltonian by the Schrieffer-Wolff transformation} \label{appsec:fragmentation}
\begin{figure}[h]
\subfigure[]{\includegraphics[width = 0.237\textwidth]{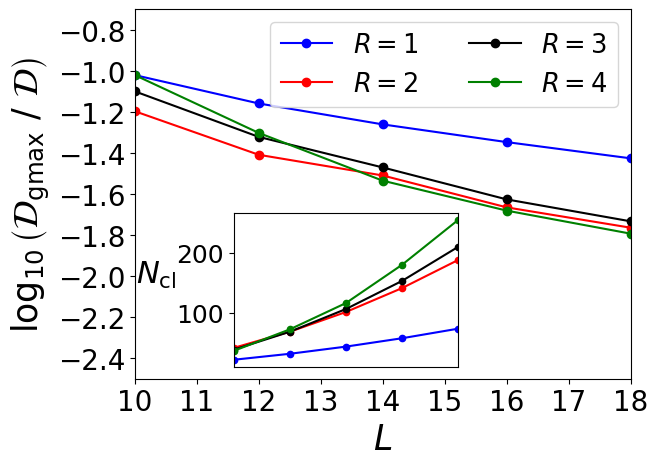}}
\subfigure[]{\includegraphics[width = 0.237\textwidth]{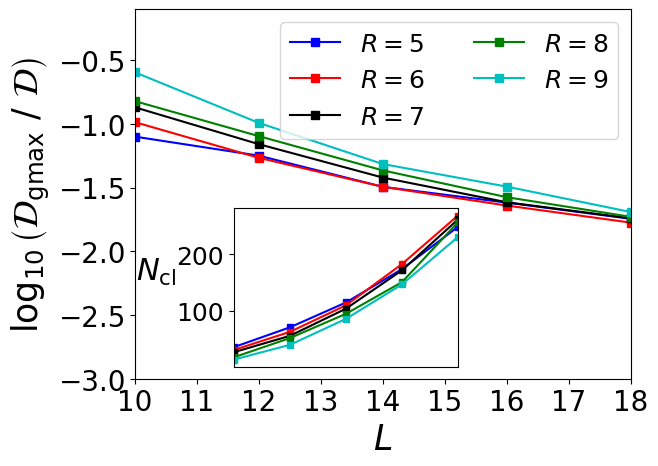}}\\
\subfigure[]{\includegraphics[width = 0.237\textwidth]{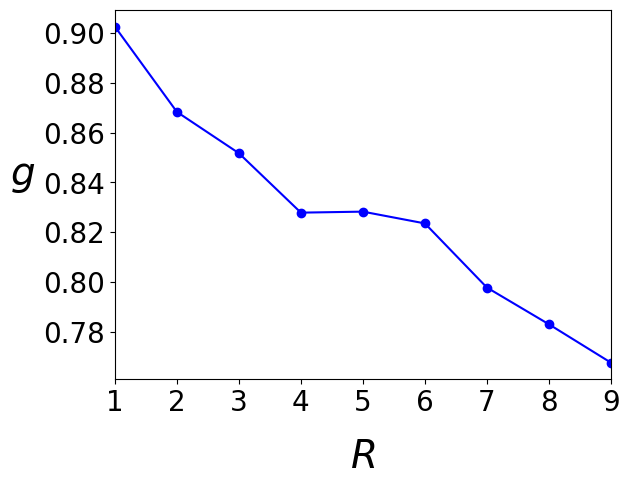}}
\caption{(a),(b) The ratio of dimension of the largest degenerate cluster = $\md_\mathrm{gmax}$ and Hilbert space dimension $\md$ [Eq.~\eqref{eq:hil_dim_clean}] as a function of system size $L$ and interaction range $R$ in log-linear scale.  The insets indicate the total number of clusters = $N_\mathrm{cl}$ as functions of $L$ for various $R$ with the same color lines as for the main plots. (c) Gradient $g$ as a function of $R$, obtained by the straight line fits on the main plots in subfigures (a) and (b) for the last four points \{$L = 12, 14, 16, 18$\} (to avoid the finite-size effect). It indicates the decay of $\md_\mathrm{gmax}/\md$ as $g^L$ with system size $L$ for various $R$.}
\label{appfig:deg_frag_ham}
\end{figure}

 At the limit $t/U = 0$, only the interaction part of the Hamiltonian \eqref{eq:ham_clean} survives. The corresponding eigenstates in the space of zero total magnetization and symmetric under spin-flip operator $\prod_n \sigma^x_n$ form a set of disconnected clusters \cite{Yang2020hilbert}, where each cluster consists of degenerate eigenstates (for $L \geq 6$), and different clusters are separated by large energy gaps at least of order $\sim U/2$. The Hilbert space is strongly fragmented \cite{Sala20, Khemani20, moudgalya2022quantum} as depicted in Fig.~\ref{appfig:deg_frag_ham}. We are not aware of any symmetry that can justify the generation of such separated clusters. The dimension of the largest degenerate cluster $\md_\mathrm{gmax}$ decreases exponentially with the increase of the system size $L$ compared to the total Hilbert space dimension $\md$ (after resolving the symmetries).
In the presence of perturbative hopping ($t \ll U$) we expect a change in energy eigensystem properties within each cluster, and negligible mixing between different clusters. 
For convenience, we define $\mh = \mh_R/U$ and split the interaction part and the hopping part $\mh$ = $\mh_\mathrm{hop}$ + $\mh_\mathrm{int}$ [see Eq.~\eqref{eq:ham_clean}].
For a general $R$ and $L$, the hopping term  $\mh_\mathrm{hop}$ = $-(t/U)$ $\sum_{n = 0}^{L-2}$ $(\sigma^+_n \sigma^-_{n+1}$ $+ \sigma^-_n \sigma^+_{n+1})$ does not commute with the interaction part $\mh_\mathrm{int}$. In such cases, for the degenerate subspace belonging to a cluster, the first-order $(t/U)^1$ perturbative energy vanishes. The hopping term $(\sigma^+_n \sigma^-_{n+1}$ + $\sigma^-_n \sigma^+_{n+1})$ only connects spins from nearest neighbors, while mixing or changing the energy spacing between nonperturbative degenerate states requires long-range correlated hoppings. For example, within the spin-flip symmetry and zero magnetization sector in $L$ = 10, the states $(\ket{{\tiny{\CIRCLE\Circle\Circle\CIRCLE \CIRCLE \Circle\Circle \CIRCLE \CIRCLE\Circle}}}$ +  $\ket{{\tiny{\Circle\CIRCLE\CIRCLE\Circle\Circle \CIRCLE\CIRCLE\Circle\Circle\CIRCLE}}})/\sqrt{2}$ and $(\ket{{\tiny{\Circle\Circle\CIRCLE\CIRCLE \Circle \CIRCLE \Circle \Circle\CIRCLE \CIRCLE}}}$ + $\ket{{\tiny{\CIRCLE\CIRCLE\Circle\Circle \CIRCLE \Circle \CIRCLE \CIRCLE \Circle \Circle}}})/\sqrt{2}$ are degenerate having nonperturbative energy $= -5U/4$ for $R = 1$, where ``$\tiny{\Circle}$'' indicates a spin-down state, and ``$\tiny{\CIRCLE}$'' indicates a spin-up state. They are connected by the long-ranged hopping term $\sim \sigma^+_1 \sigma^-_3 \sigma^+_6 \sigma^-_8$, where the lattice site index $n = 0, 1, \cdots$ is counted from the left in the states vectors.  
 For $L$ = 18, the states $(\ket{{\tiny{\CIRCLE\Circle\Circle\CIRCLE \CIRCLE \Circle\Circle \CIRCLE \CIRCLE \Circle\Circle \CIRCLE \CIRCLE\Circle\Circle \CIRCLE \CIRCLE\Circle}}} + $ $\ket{{\tiny{\Circle\CIRCLE \CIRCLE\Circle\Circle \CIRCLE \CIRCLE \Circle\Circle  \CIRCLE \CIRCLE \Circle\Circle\CIRCLE \CIRCLE\Circle\Circle \CIRCLE}}})/\sqrt{2}$, $(\ket{{\tiny{\CIRCLE\Circle\Circle\CIRCLE \CIRCLE \CIRCLE \Circle\Circle  \CIRCLE \Circle\Circle \CIRCLE \CIRCLE\Circle\CIRCLE \CIRCLE\Circle \Circle}}} + $ $\ket{{\tiny{\Circle\CIRCLE\CIRCLE\Circle\Circle\Circle\CIRCLE\CIRCLE\Circle\CIRCLE\CIRCLE\Circle \Circle\CIRCLE\Circle\Circle\CIRCLE\CIRCLE}}})/\sqrt{2}$ are degenerate having nonperturbative energy $= -9U/4$ for $R = 1$; they can be connected by the long-ranged hopping term $\sim \sigma^+_5 \sigma^-_7 \sigma^+_{14} \sigma^-_{16}$. It seems that we need a longer range of hopping for longer system size to connect all degenerate eigenstates within the same cluster or fragmented subspace. This motivates us to look for the higher-order $(t/U)^{k \geq 2}$ perturbative terms using the Schrieffer-Wolff (SW) transformation.
 
We will use a similar procedure used in the Supplemental
Sec.~IV of Ref.~\cite{Yang2020hilbert} or in Ref.~\cite{yang2024probing}.
The idea is to perform a unitary transformation controlled by an anti-Hermitian operator $\mathbb{S}$ which rotates the eigenstates perturbatively so that the resulting effective Hamiltonian becomes block-diagonal in the fragmented subspaces,   
\begin{align}
 \mh_\mathrm{eff} = &~ e^\mathbb{S} \cdot \mh \cdot e^{-\mathbb{S}} \notag\\
 = &~ \mh + [\mathbb{S}, \mh]  + \frac{1}{2}[\mathbb{S},[\mathbb{S},\mh]] + \cdots \notag\\
 = &~ \mh_\mathrm{int} + \sum_{k = 1}^{\infty} \left(t/U\right)^k \mh^{(k)}_\mathrm{eff}\;.
\end{align}
The effective Hamiltonian $\mh_\mathrm{eff}$ will be derived such that all of its terms up to a finite $k$th order $(t/U)^k$  will commute with the nonperturbative Hamiltonian $\mh_\mathrm{int}$, and $\mh_\mathrm{eff}$ obeys the spin-flip symmetry and zero magnetization conservation condition. Therefore, the derived perturbative Hamiltonian can change energy spacings between eigenstates within the same cluster only. 
It is also possible that up to a leading order perturbation each cluster further divides in many subclusters, which can give rise to another type of fragmentation. 
 
We also split the operator $\mathbb{S}$ in perturbation series
\begin{align}
 \mathbb{S} = \sum_{k = 1}^{\infty} (t/U)^k \mathbb{S}^{(k)}\;.
\end{align} 
The construction of $\mathbb{S}$ relies on the demand that all the off-diagonal terms that connect different clusters will cancel out at each perturbative order---see details in supplementary material of Ref.~\cite{Yang2020hilbert}. At first order this means
\begin{align}
\mh^{nc}_\mathrm{hop} + [\mathbb{S}^{(1)}, \mh_\mathrm{int}] = 0, \label{appeq:s_cond}
\end{align}  
where we split the hopping part in the commutative and noncommutative parts $\mh_\mathrm{hop} = (t/U)(\mh^{nc}_\mathrm{hop} + \mh^{c}_\mathrm{hop})$ such that $[\mh^{nc}_\mathrm{hop}, \mh_\mathrm{int}] \neq 0$ and   $[\mh^{c}_\mathrm{hop}, \mh_\mathrm{int}] = 0$. For example, $\mh^{c}_\mathrm{hop}$ can be like the folded hopping terms, which can induce further fragmentation within each cluster [cf.~Eq.~(7) in Ref.~\cite{yang2024probing}]. Up to second order, the effective Hamiltonian terms are 
\begin{align}
 &\mh^{(1)}_\mathrm{eff} = \mh^{c}_\mathrm{hop}, \notag\\&
 \mh^{(2)}_\mathrm{eff} = \left[\mathbb{S}^{(1)}, \mh^{c}_\mathrm{hop} + (1/2)\mh^{nc}_\mathrm{hop}\right]  + [\mathbb{S}^{(2)}, \mh_\mathrm{int}],
 \label{appeq:eff_ham}
\end{align}
where $\mathbb{S}^{(2)}$ will be derived based on the demand that $[\mathbb{S}^{(2)}, \mh_\mathrm{int}]$ will remove the irrelevant parts (which are noncommutative with $\mh_\mathrm{int}$) from the other term in $\mh^{(2)}_\mathrm{eff}$. 
 Since we are outlining the procedure, we will provide only $\mathbb{S}^{(1)}$ for the shortest range $R = 1$.
 For our $R = 1$ case,          
\begin{align}
  \mh^{c}_\mathrm{hop} = 0, \mh_\mathrm{int} =   \sum_{m=0}^{L-2} (m+1)\sigma^z_m \sigma^z_{m+1}\;. \label{appeq:int_hamil_1}
\end{align} 
To derive the form of SW operator $\mathbb{S}^{(1)}$, we take motivation from the form given in Ref.~\cite{yang2024probing}. We define projectors on the spin-down and spin-up states at each lattice site $n$,
\begin{align}
P_n = \ketbra{\small{\Circle}}{\small{\Circle}}
= \frac{1}{2} - \sigma^z_n,
~Q_n =\ketbra{\small{\CIRCLE}}{\small{\CIRCLE}}
= \frac{1}{2} + \sigma^z_n\;.                                                                                                                                                                                                                      
                                                                                                                                                                                           \end{align}
 The ansatz $\mathbb{S}^{(1)} \sim A_{n-1} \sigma^+_n \sigma^-_{n+1} B_{n+2}$ is taken where $A, B$ is one of the projectors $P, Q$. 
 For convenience, we define 
 \begin{align} A^n_{pp} = P_{n-1} \sigma^+_n \sigma^-_{n+1} P_{n+2}\;.\end{align} 
Similar definitions follow for $A^n_{pq}, A^n_{qp}$, and $A^n_{qq}$.  The following commutation relations will be used to derive $\mathbb{S}^{(1)}$:    
\begin{align}
& [A^n_{pp}, (m+1)\sigma^z_m \sigma^z_{m+1}] \notag\\
& = \delta_{m,n-1} n (-P_{n-1}/2)(-\sigma^+_n) \sigma^-_{n+1} P_{n+2} \notag\\
& + \delta_{m,n+1} (n+2) P_{n-1} \sigma^+_n \sigma^-_{n+1} (-P_{n+2}/2)\notag\\
& \implies \sum_{n=0}^{L-2} \sum_{m=0}^{L-1}\left[ A^n_{pp},  (m+1)\sigma^z_m \sigma^z_{m+1}\right]  \notag\\
& = -\sum_{n=0}^{L-2}  P_{n-1} \sigma^+_n \sigma^-_{n+1} P_{n+2}\;.\label{appeq:pp_1}
\end{align} 
 Note that the original interaction term Eq.~\eqref{appeq:int_hamil_1} does not carry the term $\sigma^z_{L-1} \sigma^z_{L}$, which we added in the above commutation relation to obtain an identical coefficient = $-1$ for all $n$ in the final result of the commutator. Including this extra term $\sigma^z_{L-1} \sigma^z_{L}$ in Eq.~\eqref{appeq:int_hamil_1} will not affect the dynamics if we staple the spin configurations at the last two sites $n = L-1, L$ so that the spin at extra added site $n = L$ takes a fixed orientation depending on the spin state at the previous site, either $\ket{\small{\CIRCLE \CIRCLE}}_{L-1,L}$ or $\ket{\small{\Circle \Circle}}_{L-1,L}$, it will just add a constant term in the main Hamiltonian. Without this extra term, it is also possible to obtain the commutator. As in Ref.~\cite{yang2024probing}, we consider $P$, $Q$ operators at the added sites $n= -1, L$, which will not influence the dynamics. Henceforth we will include these considerations, 
 \begin{align}
 &[A^n_{pq}, (m+1)\sigma^z_m \sigma^z_{m+1}] \notag\\
 & = \delta_{m,n-1} n (-P_{n-1}/2)(-\sigma^+_n) \sigma^-_{n+1} Q_{n+2} \notag\\
 & + \delta_{m,n+1} (n+2) P_{n-1} \sigma^+_n \sigma^-_{n+1} (Q_{n+2}/2)\notag\\
 & \implies \sum_{n=0}^{L-2} \sum_{m=0}^{L-1}\left[A^n_{pq},  (m+1)\sigma^z_m \sigma^z_{m+1}\right]  \notag\\
& = \sum_{n=0}^{L-2} (n+1) P_{n-1} \sigma^+_n \sigma^-_{n+1} Q_{n+2},\label{appeq:pq_1}
\end{align}
\begin{align}
 &[A^n_{qp}, (m+1)\sigma^z_m \sigma^z_{m+1}] \notag\\
 & = \delta_{m,n-1} n (Q_{n-1}/2)(-\sigma^+_n) \sigma^-_{n+1} P_{n+2} \notag\\
 & + \delta_{m,n+1} (n+2) Q_{n-1} \sigma^+_n \sigma^-_{n+1} (-P_{n+2}/2)\notag\\
 & \implies \sum_{n=0}^{L-2} \sum_{m=0}^{L-1}\left[A^n_{qp},  (m+1)\sigma^z_m \sigma^z_{m+1}\right]  \notag\\
& = -\sum_{n=0}^{L-2} (n+1) Q_{n-1} \sigma^+_n \sigma^-_{n+1} P_{n+2},\label{appeq:qp_1}
\end{align}
\begin{align}
 & [A^n_{qq}, (m+1)\sigma^z_m \sigma^z_{m+1}] \notag\\
 & = \delta_{m,n-1} n (Q_{n-1}/2)(-\sigma^+_n) \sigma^-_{n+1} Q_{n+2} \notag\\
 & + \delta_{m,n+1} (n+2) Q_{n-1} \sigma^+_n \sigma^-_{n+1} (Q_{n+2}/2)\notag\\
 & \implies \sum_{n=0}^{L-2} \sum_{m=0}^{L-1}\left[A^n_{qq},  (m+1)\sigma^z_m \sigma^z_{m+1}\right]  \notag\\
 & = \sum_{n=0}^{L-2}  Q_{n-1} \sigma^+_n \sigma^-_{n+1} Q_{n+2}\;.\label{appeq:qq_1}
\end{align} 
From Eqs.~\eqref{appeq:pp_1}--\eqref{appeq:qq_1} we obtain 
 \begin{align}
   &\sum_{n=0}^{L-2} \sum_{m=0}^{L-1} \Bigg[- A^n_{pp} + A^n_{qq} + \frac{A^n_{pq} - A^n_{qp}}{n+1} , ~ (m+1)\sigma^z_m \sigma^z_{m+1}\Bigg]\notag\\ &
  = \sum_{n=0}^{L-2}  (P_{n-1} + Q_{n-1}) \sigma^+_n \sigma^-_{n+1} (P_{n+2} +  Q_{n+2}) \notag\\&
  = \sum_{n=0}^{L-2} \sigma^+_n \sigma^-_{n+1}\;.
\end{align} 
 Therefore, in order to satisfy Eq.~\eqref{appeq:s_cond}, we need 
\begin{align}
 \mathbb{S}^{(1)} =  \sum_{n=0}^{L-2}
 \Bigg[P_{n-1} (\sigma^+_n \sigma^-_{n+1} -  \sigma^-_n \sigma^+_{n+1}) \left(\frac{Q_{n+2}}{n+1} - P_{n+2}\right) \notag\\
 + Q_{n-1} (\sigma^+_n \sigma^-_{n+1} -  \sigma^-_n \sigma^+_{n+1})\left( Q_{n+2} - \frac{P_{n+2}}{n+1} \right)\Bigg]\;. \label{appeq:SW_S1}
\end{align}
The $\mathbb{S}^{(1)}$ carries nearest-neighbor hopping terms with projector constraints that are separated by three lattice unit cells. Therefore, in $\mh^{(2)}_\mathrm{eff}$, which has a single $\mathbb{S}^{(1)}$ operator in the commutator [Eq.~\eqref{appeq:eff_ham}], we can perform correlated hopping $\sim$ $\sigma^+_n \sigma^-_{n+1} \sigma^+_{n+2} \sigma^-_{n+3}$ between sites separated  maximally by three unit cells. Similarly, $\mh^{(3)}_\mathrm{eff}$, which has a maximum three $\mathbb{S}^{(1)}$ operators in the commutator (cf.~the Supplemental Material of Ref.~\cite{Yang2020hilbert}), can perform hopping between different lattice sites which are separated by seven lattice unit cells. Therefore, we expect the $\mh^{(3)}_\mathrm{eff}$ will connect the example states  $(\ket{{\tiny{\CIRCLE\Circle\Circle\CIRCLE \CIRCLE \Circle\Circle \CIRCLE \CIRCLE\Circle}}}$ +  $\ket{{\tiny{\Circle\CIRCLE\CIRCLE\Circle\Circle \CIRCLE\CIRCLE\Circle\Circle\CIRCLE}}})/\sqrt{2}$ and $(\ket{{\tiny{\Circle\Circle\CIRCLE\CIRCLE \Circle \CIRCLE \Circle \Circle\CIRCLE \CIRCLE}}}$ + $\ket{{\tiny{\CIRCLE\CIRCLE\Circle\Circle \CIRCLE \Circle \CIRCLE \CIRCLE \Circle \Circle}}})/\sqrt{2}$ discussed at the beginning of this section. If $\mh^{(3)}_\mathrm{eff}$ cannot connect all the states within a cluster, we can check the next higher-order perturbed effective Hamiltonian, and the iteration can be performed until we obtain the leading-order correction that will fulfill our demands.  In general, $\mh^{(k)}_\mathrm{eff}$ involves commutators where $\mathbb{S}^{(1)}$ participates maximally $k$-number ($k$ $>$ $2$) of times, which implies $\mh^{(k)}_\mathrm{eff}$ can carry hopping among lattice sites separated by $2k+1$ lattice units $\sim \sigma^+_n \cdots \sigma^-_{n + 2k + 1}$. Finally, the effective Hamiltonian $\mh_\mathrm{eff}$ can carry both diagonal and off-diagonal matrix elements with respect to the nonperturbative degenerate eigenbasis, and $\mh_\mathrm{eff}$ mixes different eigenstates within an individual cluster. As a result, the Krylov space (related to each of the fragmented Hilbert subspaces) restricted thermalization can arise \cite{yang2024probing}.

Following a similar procedure as in Eqs.~\eqref{appeq:pp_1}--\eqref{appeq:SW_S1}, one can derive the expression for the full SW operator $\mathbb{S}$ for arbitrary $R$. For all the cases, we expect mixing among states within a cluster only because of resonance, and no mixing among states from different clusters because of off-resonance (large energy gaps). Hence the fragmentation structure of Hilbert space will be well maintained for perturbative $t/U$.  

\bibliography{tilt_spin_dis}

\end{document}